%% file: bttf-arxiv.tex
\def\ARXIVVersion{1}
\documentclass[acmsmall,screen,nonacm]{acmart}

\usepackage{mathtools}
\usepackage{graphicx}
\usepackage{booktabs}
\usepackage{array}
\usepackage{colortbl}
\usepackage{float}
\usepackage{multirow}
\usepackage{needspace}
\usepackage[most]{tcolorbox}
\usepackage[nameinlink]{cleveref}

\newcommand{\ourModel}{BTTF}
\definecolor{cloudllm}{RGB}{78,121,167}
\definecolor{cloudllmbg}{RGB}{239,244,250}
\newcommand{\suppref}[1]{%
  \ifdefined\FSEMainOnly
    the supplementary material%
  \else
    \Cref{#1}%
  \fi}
\newcommand{\mainpaperref}[2]{\cref{#2}}

\newtcolorbox{findingbox}[1]{
  enhanced,
  colback=cyan!4,
  colframe=cyan!45!black,
  colbacktitle=cyan!14,
  coltitle=black,
  fonttitle=\bfseries,
  title={#1},
  boxrule=0.6pt,
  arc=1mm,
  left=1.5mm,
  right=1.5mm,
  top=1mm,
  bottom=1mm,
  before skip=6pt,
  after skip=6pt
}

\AtBeginDocument{%
  }

\setcopyright{none}
\title[Back to the Future: Regressing Readability Features from LLMs]
{Back to the Future: Regressing Readability Features\\ from LLMs}
\author{Tairan Wang}
\ifdefined\ARXIVVersion
\affiliation{%
  \institution{University College London}
  \city{London}
  \country{United Kingdom}}
\email{tairan.wang.22@ucl.ac.uk}
\fi
\author{Earl T. Barr}
\ifdefined\ARXIVVersion
\affiliation{%
  \institution{University College London}
  \city{London}
  \country{United Kingdom}}
\email{e.barr@ucl.ac.uk}
\fi

\ifdefined\FSEMainOnly
\fi

\begin{document}

\input{Sections/abs.tex}

\maketitle

\input{Sections/intro.tex}
\input{Sections/method.tex}
\input{Sections/eval.tex}
\input{Sections/results.tex}

\input{Sections/disc.tex}
\input{Sections/relwork.tex}
\input{Sections/conc.tex}
\input{Sections/data_availability.tex}

\bibliographystyle{ACM-Reference-Format}
\bibliography{references}

\ifdefined\FSEMainOnly
\else
\appendix
\crefalias{section}{appendix}
\crefalias{subsection}{appendix}
\input{Appendices/features.tex}
\input{Appendices/feature_sele.tex}
\input{Appendices/reada_dataset.tex}
\input{Appendices/baseline.tex}
\input{Appendices/degr_dataset.tex}
\input{Appendices/ablation.tex}
\input{Appendices/robustness.tex}
\fi

\end{document}

%% file: Sections/abs.tex
\begin{abstract}

Code readability determines how effectively and efficiently software can be
understood, reviewed, and maintained. As AI coding agents assume a larger role
in software engineering, readability gains a critical dual purpose: it allows
human engineers to safely review agent-generated output, and it helps the
agents themselves maintain code more effectively, as readable code tends to be
more compact and less distracting within limited context windows. Yet, existing
readability models do not consistently agree with human judgements across
datasets.
 
To address this, we introduce \ourModel{} (\textbf{B}ack \textbf{T}o \textbf{T}he
\textbf{F}uture), a model built on the principle of pairing an old-school
linear regression backbone with new-school information-theoretic features.
Rather than using large language models as opaque, black-box judges, \ourModel{}
leverages them as measurement instruments. \ourModel{} combines
traditional code measurements with embedding-based semantic organisation and
causal-LM predictability to capture source structure, semantic coherence, and
contextual surprise.
 
We evaluate \ourModel{} across the complete set of five publicly available
human-rated code-readability benchmarks, supplemented by our manually annotated
MBJP development set, for a total of 1,100 samples. Across this comprehensive
suite, \ourModel{} establishes a new state of the art, improving
average Spearman correlation over the strongest baseline by up to 0.101. Across
13 controlled readability-reducing transformations applied to production code,
it achieves response rates of 93.8\% in Java and 87.6\% in Python,
outperforming Dorn by 14\% and 19.2\%. Its performance is the highest among
local models and trails the strongest cloud-LLM response by 1.7\% in Java
and 5.1\% in Python.
The direct-LLM baselines require an average of 1,083 prompt tokens per evaluation
while remaining opaque. Unlike alternative methods, \ourModel{}
delivers state-of-the-art benchmark performance and strong transformation
robustness alongside feature-granular interpretability, with every feature's
coefficient strictly matching its theoretical directional effect across all
evaluated model configurations. 

\end{abstract}

%% file: Sections/intro.tex
\section{Introduction}
\label{sec:introduction}

Code readability is central to software engineering because source code is
repeatedly read, reviewed, debugged, and modified throughout its lifetime. The
need for reliable automated assessment has now become urgent as LLM-driven
coding agents rapidly proliferate. As agents shift human engineers from primary
authors into reviewers, high throughput makes human inspection the central
bottleneck: a strain compounded by legal and regulatory mandates that require
human auditing of deployed software for the foreseeable future, even if source
code eventually becomes the ``new assembly''. In this emerging paradigm,
readability serves a critical dual purpose: it eases the burden on human
reviewers evaluating agent-generated output, and it enables agents themselves
to navigate and reason over codebases more effectively without cluttering
context windows. Readability must therefore be measured explicitly,
interpretably, and at scale to give developers actionable feedback, guide LLM
post-training, and enable agents to optimise code for both human auditing and
downstream maintenance. Despite many proposed measures and predictive models,
aligning automated metrics with human judgement
remains an open challenge.

Existing work follows three main directions. Feature-based machine-learning
models use explicit lexical, structural, textual, and visual measurements
\cite{buse2009model,posnett2011model,dorn2012model,
scalabrino2018model}. Smaller formulations rely on broad proxies, while
larger feature inventories can combine language- or presentation-specific
measurements whose fitted directions are difficult to interpret
consistently. Deep-learning models instead learn representations directly
from source code, but provide less visibility into the properties that
determine a prediction \cite{mi2018model1,mi2022model2}. More recently,
language models have been used to generate readability judgements or to
measure whole-program prediction difficulty, although their reported
agreement with human ratings remains variable
\cite{ouedraogo2025llm_method,dillmann2024llm_prob_signal_method}.
Across these directions, the two core challenges remain:
generalisation and explainability.

\begin{figure*}[t]
    \centering
    \includegraphics[width=\textwidth]
    {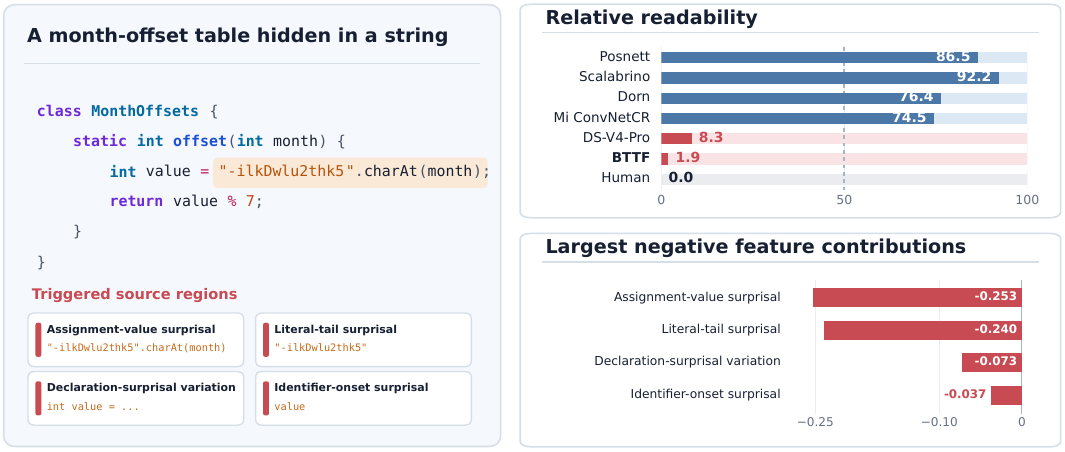}
    \caption{A well-known compact day-of-week implementation that encodes
    month offsets as a string. All five developers assign it the minimum
    rating of 1/5, mapped to 0 on the common 0--100 scale. Model scores are
    percentiles over the 1,100 human-rated samples; lower values indicate
    lower readability. DS-V4-Pro is averaged over five judgements. The
    lower-right panel shows \ourModel's four largest negative feature
    contributions.}
    \label{fig:motivating-example}
\end{figure*}

Figure~\ref{fig:motivating-example} makes the first challenge concrete. The
example hides a month-offset lookup inside a string and character-code
calculation. All five developers who rated the example assign it the minimum
readability score of 1/5. Four established readability models nevertheless
place it between the 74th and 92nd percentiles of their benchmark predictions.
DS-V4-Pro and \ourModel{} both recognise the code as difficult, but
\ourModel{} also identifies the specific source regions and measurements
responsible for its low score.

People do not understand complex material by retaining every detail
independently. Related and recurring elements can instead be represented
through shared patterns and higher-level abstractions. From an
information-theoretic perspective, this is compression: many observations are
encoded by a shorter internal description
\cite{chater2003simplicity,feldman2016simplicity}. Program comprehension
likewise produces mental representations at multiple levels of abstraction
\cite{pennington1987mental_representation,
bidlake2020mental_representations}. In source code, identifiers, expressions,
assignments, and control structures may collectively express concepts such as
input validation, state updates, or iteration.

To address these challenges, we introduce \ourModel{} (\textbf{B}ack \textbf{T}o
\textbf{T}he \textbf{F}uture), a readability model built on the principle of
combining an old-school linear regression backbone with new-school
information-theoretic features. Rather than treating large language models as
opaque, end-to-end judges, \ourModel{} uses pretrained causal LMs and continuous
embedding spaces purely as measurement instruments. Linear regression provides
transparency by ensuring that every feature's contribution is
explicitly weight-attributed and directionally verifiable. Meanwhile, the
underlying features capture source structure, contextual surprise,
and semantic coherence.

To measure semantic coherence, \ourModel{} leverages pretrained embedding spaces as
geometric proxies for mental consolidation. Continuous embedding models place
semantically related inputs near one another in representation space
\cite{nussbaum2024nomicembedding,zhang2025qwen3embedding,gunther2023jinaembedding,yu2024snowembedding}.
Coherent groups of identifiers and control structures reflect information that
is readily consolidated by a reader, whereas fragmented distributions indicate
elements that are more difficult to organise into common abstractions.

Embedding spaces capture which program elements are related, but
not whether they appear where a reader would expect them, as, for example when
unrelated logic is interleaved with the current computation. A name may be
unexpected when first encountered, an assignment value may not follow from
preceding statements, or a short identifier may require context from much
earlier in the program. Causal code language models complement semantic
organisation by measuring the prediction difficulty of observed source
constructs, revealing which elements are unexpected and whether additional
preceding context makes them easier to anticipate. Together, the measurements
describe how source information forms semantic units, unfolds through the
program, and relates to its preceding context.

Based on this view, we use pretrained models as measurement instruments rather
than as end-to-end readability predictors. Stability analysis and consensus
ranking across three causal language models and five embedding models select
features that remain useful under different pretrained representations. The
resulting model contains 11 named, directly interpretable features: four
traditional measurements of source complexity, two measurements of semantic
organisation in embedding space, and five measurements of contextual prediction
difficulty derived from causal code language models
(\Cref{tab:final-feature-schema} defines all 11).

We evaluate \ourModel{} across the complete set of five publicly available human-rated
code-readability benchmarks, supplemented by our manually annotated MBJP
development set, for a total of 1,100 samples. Across this comprehensive suite,
\ourModel{} establishes a new state of the art under both pooled cross-validation
and leave-one-dataset-out (LODO) evaluation, achieving average Spearman correlations
of 0.631 and 0.638—gains of 0.094 and 0.101 over the strongest baseline average
(paired 95\% bootstrap CIs [0.031, 0.157] and [0.038, 0.165]; $p \le 0.0032$). Across 13
controlled readability-reducing transformations applied to production code,
\ourModel{} achieves expected-direction response rates of 93.8\% in Java and 87.6\% in
Python, outperforming Dorn by 14\% and 19.2\%. It assigns lower readability to a
majority of changed variants for every transformation in both languages. While
approaching the robustness of opaque frontier models (falling within 1.7 and 5.1
percentage points of GPT-6.1 Sol without incurring remote generative overhead
averaging 1,083 prompt tokens per evaluation),
\ourModel{} provides feature-granular interpretability via signed
contributions from 11 source properties,
where every fitted coefficient strictly matches its theoretical directional effect across
all evaluated model configurations (\Cref{tab:coefficient-stability}).

This work makes three key contributions:

\begin{itemize}

    \item We introduce \ourModel, an interpretable, linear regression model for
        readability that combines traditional code measurements with
        semantic organisation and information-theoretic predictability derived
        from pretrained models (\Cref{sec:approach}).

    \item We show that \ourModel{} achieves state-of-the-art Spearman correlations
        of 0.631 (pooled) and 0.638 (LODO), improving baseline averages by 0.094
        and 0.101. It approaches GPT-6.1 Sol response rates with 93.8\% (Java) and 87.6\%
        (Python) transformation response rates without remote prompting or response
        generation, while
        ensuring fully interpretable feature contributions
        (\Cref{tab:benchmark-results,tab:interference-results}).

    \item We deliver unmatched interpretability: every feature in \ourModel{} aligns
        with its theoretical effect without sign reversals across configurations
        (\Cref{sec:model-explainability,tab:coefficient-stability}). In contrast,
        frontier LLMs such as GPT-6.1 overrate subtle code defects by relying on
        high-level summaries that fail to properly weigh combined property severities
        (\Cref{sec:model-explainability}).

\end{itemize}

%% file: Sections/method.tex
\section{Back to the Future: Interpretable Code Readability using LLM-Derived Signals}
\label{sec:approach}

\ourModel{} models readability along two information dimensions: how
compactly program elements organise into semantic abstractions, and how
predictably information unfolds through the source. For the first, we extract
AST-tagged code snippets and embed them individually. Geometric and clustering
measurements then distinguish code whose identifiers and structures form a
small number of coherent semantic groups from code whose elements remain
dispersed. Such organisation reflects information compression: many
source-level observations can be represented through fewer higher-level
patterns.

For the second dimension, we use frozen causal code language models to measure
the surprisal of observed source constructs. Rather than reduce an entire
program to a single perplexity value, we locate where new information arises,
which kinds of constructs are difficult to predict, and whether additional
preceding context reduces that difficulty. These measurements capture the
ordering and contextual support that embedding-space organisation alone
cannot provide. The complete representation combines these two dimensions
with traditional measurements of surface-level information properties.

\subsection{Feature Construction and Selection}

\begin{figure*}[t]
\centering
\includegraphics[width=\textwidth]{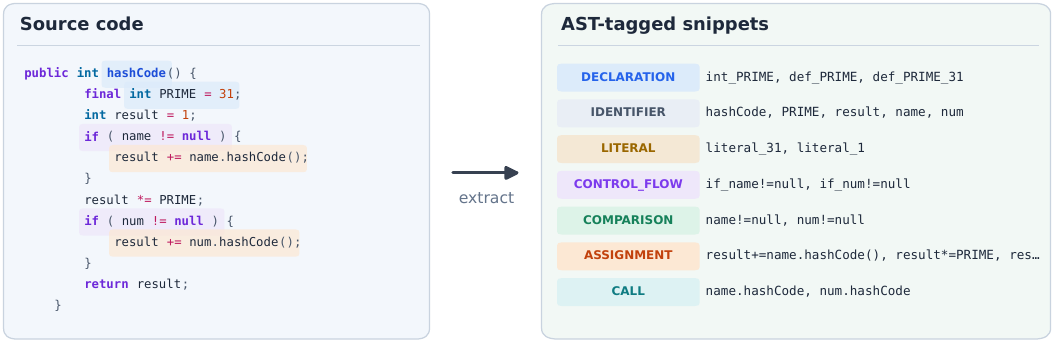}
\caption{AST-tagged snippet representation of a Java method from the Scalabrino
dataset. Source constructs are converted into normalised strings while
preserving program roles and multiple structural levels; the right panel shows
representative strings grouped by AST tag.}
\label{fig:AST-tagged-snippet-extraction}
\end{figure*}

Embedding models operate on strings rather than program structures. We
therefore extract source constructs as AST-tagged snippets and convert them into
normalised representations. Raw source strings can mislead an
embedding model: surface differences may obscure related program information,
while compound expressions may blur distinct elements. We address these cases
in opposite ways: compound constructs are decomposed so that their internal
elements remain distinguishable, whereas elements that share a program role,
such as literals, receive a common textual cue while retaining their
original content. The resulting snippets provide the inputs for
semantic-organisation measurements.
We also compute traditional features directly from the source and extracted
snippets to capture surface-level information properties.
The resulting representation is illustrated in~\Cref{fig:AST-tagged-snippet-extraction}.

Embeddings allow us to compare the semantic similarity of code snippets. We
use these similarities for automatic clustering, which shows whether the snippets
concentrate into a few coherent semantic groups or remain fragmented.
The former reflects a compact higher-level organisation of program information;
the latter reflects a larger set of distinct abstractions. To avoid relying on
a single grouping assumption, we use multiple adaptive clustering methods and
retain geometric measurements as complementary evidence.

\begin{figure*}[t]
    \centering
    \includegraphics[
        width=\textwidth,
        trim={0 0 0 0},
        clip
    ]{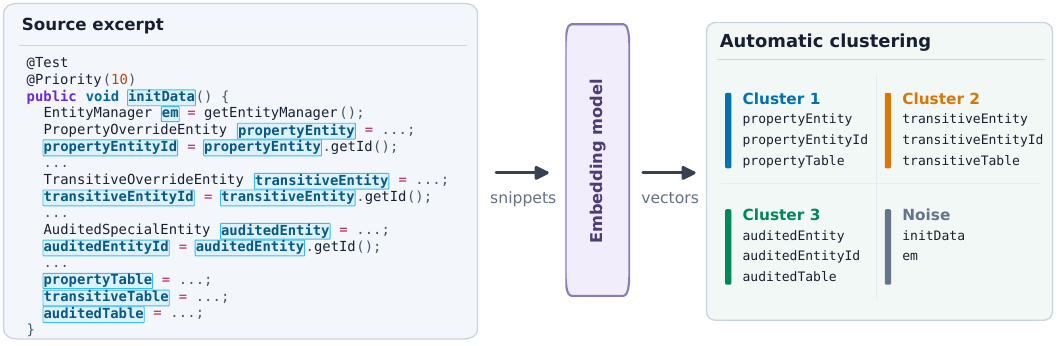}
    \caption{Embedding-space organisation of identifiers extracted
    from a Java method in the Scalabrino dataset. Identifier strings are
    mapped to vectors by a pretrained embedding model and grouped using
    OPTICS; colours indicate cluster assignments, and grey indicates noise.}
    \label{fig:snippet-clustering-motivation}
\end{figure*}

This process is illustrated using a real code sample
in~\Cref{fig:snippet-clustering-motivation}. Identifiers are extracted,
embedded, and grouped
automatically, allowing semantically related names to form coherent families
while keeping unrelated identifiers separate.

A causal code language model provides a complementary measure of how well each
observed element fits the preceding source. For token $t_i$, this consistency is
quantified by surprisal~\cite{levy2008surprisal}:

\[
s_i=-\log_2 p(t_i\mid t_{<i}).
\]

Lower values indicate stronger consistency with the preceding context.
Prior work shows why a single aggregate is insufficient: whole-class
cross-entropy alone is less effective than logical lines of code (LLOC) for
predicting maintainability-related
judgements~\cite{dillmann2024llm_prob_signal_method}. We therefore align token
surprisal with AST-tagged source constructs and compute separate measurements for
identifiers, literals, declarations, assignments, and other construct AST tags.
We also compare the same targets under shorter and longer
preceding contexts, revealing how much earlier source reduces their surprise.
Complete feature definitions, model lists,
normalisation, and calibration procedures appear in~\suppref{app:features}.

We first exclude candidates whose definitions are not portable or do not
support clear interpretation, preventing them from influencing the selection
of valid measurements. The remaining candidates are evaluated on the pooled
six datasets separately under each pretrained-model configuration. We then
combine the rankings of the same feature across configurations, favouring
measurements that remain stable when the pretrained models change. \ourModel{}
retains the literal top 11 positions of this consensus ranking. Across all six
datasets, adding lower-ranked features beyond these 11 produces only small,
dataset-dependent changes in performance. Complete eligibility rules, ranking
procedures, feature budgets, and
candidate audits appear in~\suppref{app:selection}. For comparison, we
also retain the independently selected 18-feature embedding-only model
developed before causal-LM measurements were introduced; it is a historical
predecessor rather than a post-hoc ablation of \ourModel.

\subsection{Final \ourModel{} Model}

The final \ourModel{} model uses 11 features: four traditional
measurements, two embedding-derived measurements, and five causal-LM
predictability measurements. They are summarised in~\Cref{tab:final-feature-schema}.
All features are oriented towards increasing information or
reading demand. Complete mathematical definitions are provided
in~\suppref{app:features}; the fitted directions and their stability across
pretrained models are examined in~\cref{sec:results}.

\begin{table*}[t]
\centering
\small
\setlength{\tabcolsep}{3.5pt}
\caption{Final feature schema. BPB denotes bits per source byte
(\Cref{eq:BPB}); larger values indicate greater causal-LM prediction
difficulty.}
\label{tab:final-feature-schema}
\begin{tabular}{p{0.15\textwidth}p{0.29\textwidth}p{0.50\textwidth}}
\toprule
\textbf{Family} & \textbf{Feature} & \textbf{Interpretation} \\
\midrule
Traditional
& Operator density
& Operator tokens per non-empty source line. \\

Traditional
& Decision density
& Comparison operators per non-empty source line. \\

Traditional
& Expression--literal density
& Compound expression structure relative to explicitly stated values. \\

Traditional
& Longest line length
& Maximum length of a non-empty source line. \\

Embedding
& Computation-and-control pattern count
& Number of semantic groups formed by assignment, operator, and control-flow
constructs. \\

Embedding
& Identifier embedding dispersion
& Fraction of identifier-embedding variance not explained by the first
principal component. \\

Causal LM
& Literal-tail surprisal
& Mean surprisal among the highest-surprisal 20\% of literal spans. \\

Causal LM
& Identifier-onset surprisal
& Surprisal at the beginning of identifiers. \\

Causal LM
& Assignment-value surprisal
& Mean surprisal of assignment right-hand sides. \\

Causal LM
& Declaration-surprisal variation
& Variation in surprisal across declaration headers. \\

Causal LM
& Short-identifier context dependence
& Mean reduction in short-identifier BPB when longer preceding context is
available. \\
\bottomrule
\end{tabular}
\end{table*}

\paragraph{Traditional features.}
The four traditional features describe surface-level properties of expressions
and layout.
\textit{Operator density} measures how many operations a reader must parse
within each non-empty source line. From an information-theoretic
perspective, it records the concentration of symbolic transformations within
the same reading unit.

\textit{Decision density} measures the concentration of relational and
equality comparisons. Each comparison distinguishes between possible
program states, and the feature records how many such alternatives occur
within each non-empty source line.

\textit{Longest line length} records the length of the longest non-empty
line. A longer uninterrupted span presents more source information before
a line boundary provides a natural segmentation point.

\textit{Expression--literal density} contrasts compound operations and
relations with values stated directly in the source. Its exact engineering
definition appears in~\suppref{app:features}.

\paragraph{Embedding-space signals.}
The two embedding features ask whether the concepts within a program form a
small, coherent organisation or are scattered across several distinct semantic
regions.

\textit{Computation-and-control pattern count} groups control-flow, assignment,
arithmetic, comparison, logical, and bitwise snippets according to their semantic
similarity and counts the automatically identified groups. Recurring and
conventional forms are absorbed into the same groups and add little new
organisational information, whereas constructs that do not fit those groups
require additional patterns.

\textit{Identifier embedding dispersion} measures whether the identifier
meanings follow a common semantic direction or spread across several unrelated
directions. It is computed as one minus the fraction of identifier-embedding
variance explained by the first principal component, recording how broadly the
identifier meanings spread beyond a common direction.

\paragraph{Causal-LM predictability signals.}
The remaining five features measure how difficult selected program elements
are for a frozen causal code language model to predict from preceding source.
Adapting the standard bits-per-byte (BPB) normalisation used in language-model
evaluation~\cite{gao2020pile}, we report for a source region \(R\):

\begin{equation}
\label{eq:BPB}
\operatorname{BPB}(R)=
\frac{\text{token-surprisal bits assigned to }R}
     {\text{UTF-8 source bytes covered by }R}.
\end{equation}

This normalisation makes values more comparable across tokenisers: splitting
the same identifier into more tokenizer pieces does not by itself increase the
denominator. Higher BPB means that the observed source is less expected in its
context.

\begin{figure*}[t]
\centering
\includegraphics[width=\textwidth]
{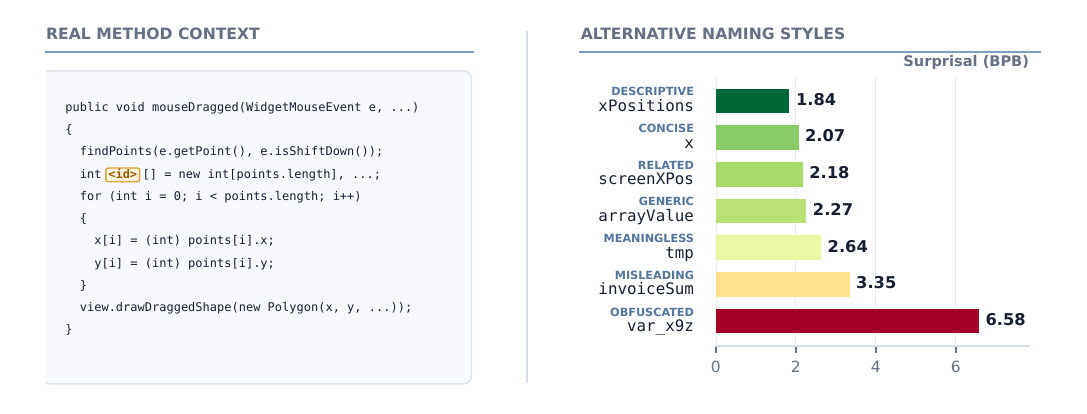}
\caption{Identifier surprisal under a fixed source context. Alternative names
replace the same variable binding, while the preceding source remains
identical and BPB is measured only at the declaration occurrence. Values are
actual measurements from Qwen2.5-Coder-0.5B; lower BPB indicates that a name
is more expected from the preceding code.}
\label{fig:identifier-surprisal}
\end{figure*}

In~\Cref{fig:identifier-surprisal}, the preceding source is held fixed and
only the variable name changes. Context-appropriate names require fewer bits
to predict, whereas vague, misleading, and obfuscated alternatives require
progressively more.

\textit{Literal-tail surprisal} first computes BPB separately for literal
occurrences and averages the most difficult 20\%. It targets exceptional
constants or encodings that may be hidden by an otherwise predictable snippet.
\textit{Identifier-onset surprisal} measures the surprisal allocated to the
first tokenizer piece of every identifier. The first piece carries the initial
lexical cue to a name and isolates its predictability at first encounter.
\textit{Assignment-value surprisal} measures the
byte-weighted prediction difficulty of initializer and assignment right-hand
sides, capturing how readily a reader can anticipate the computations that
produce stored values.

\textit{Declaration-surprisal variation} measures how consistently declarations
can be anticipated within the same source unit. We first calculate BPB
separately for each variable declaration or function signature, considering the declaration itself rather than its body or assigned value, and then
measure the variation among them. The resulting feature distinguishes uniform
declaration predictability from mixtures of conventional and unexpected forms.

Finally, \textit{short-identifier context dependence} compares each eligible
short identifier under 32-token and 256-token preceding contexts, with gain
\(g=\operatorname{BPB}_{32}-\operatorname{BPB}_{256}\). The mean gain records
how much the prediction of short identifiers relies on farther preceding code
rather than their immediate surroundings.

%% file: Sections/eval.tex
\section{Evaluation}
\label{sec:evaluation}

We use two main forms of evaluation. First, the complete set of five publicly
available human-rated code-readability benchmarks, together with our MBJP
development set, measures agreement with human judgements and transfer across
datasets. Second, controlled source-code interference experiments measure
whether predictions decrease under targeted readability reductions in Java
and Python. The following research questions specify the comparisons and
analyses conducted within these two evaluation settings.

\subsection{Research Questions}

We organise the evaluation around four research questions:

\begin{description}
    \item[\textbf{RQ1.}]
    \emph{Does \ourModel{} align more closely with human readability
    judgements than existing models do?}

    \item[\textbf{RQ2.}]
    \emph{Does \ourModel{} respond more consistently to controlled
    readability-reducing transformations than existing models do?}

    \item[\textbf{RQ3.}]
    \emph{Are \ourModel{}'s results more stable?}

    \item[\textbf{RQ4.}]
    \emph{How interpretable are \ourModel{}'s results?}
\end{description}

\subsection{Comprehensive Literature-Derived Benchmark Suite}

We aggregate the complete set of human labelled code-readability and
cognitive-complexity benchmarks established in the software engineering
literature: Buse, Scalabrino, Dorn, Schnappinger, and JetBrains. To our
knowledge, they represent the total corpus of public benchmarks evaluated
across prior literature on this topic. We supplement this literature-wide set
with our manually labelled MBJP development set, yielding a unified benchmark
of 1,100 human-rated code snippets. The suite spans Java, Python, and CUDA code
at snippet, method, class, file, and program granularity, with targets
derived from mean label values, posterior expectations, or readable-vote fractions.
The six datasets are summarised in~\Cref{tab:dataset-summary}. Detailed target
construction, annotation statistics, and the MBJP development-set procedure are
reported in~\suppref{app:datasets}.

\begin{table*}[t]
\centering
\small
\setlength{\tabcolsep}{4pt}
\caption{Summary of the human-rated readability datasets.}
\label{tab:dataset-summary}
\begin{tabular}{lrlp{0.12\textwidth}p{0.33\textwidth}}
\toprule
\textbf{Dataset}
& \textbf{Samples}
& \textbf{Language}
& \textbf{Granularity}
& \textbf{Readability Score} \\
\midrule
MBJP
& 17 & Java & Program & Mean of two ratings (1--5) \\
Buse~\cite{buse2009model}
& 100 & Java & Snippet & Mean of 121 ratings (1--5) \\
Scalabrino~\cite{scalabrino2018model}
& 200 & Java & Method & Mean of nine ratings (1--5) \\
Dorn~\cite{dorn2012model}
& 360 & Java/Python/CUDA & Snippet & Mean human ratings (1--5) \\
Schnappinger~\cite{schnappinger2020dataset}
& 304 & Java & Class & Expected readability (1--4) \\
JetBrains~\cite{sergeyuk2024dataset}
& 119 & Java & Snippet & Fraction of readable votes in $[0,1]$ \\
\bottomrule
\end{tabular}
\end{table*}

\subsection{Experimental Setup}

We evaluate the final 11-feature representation across six
established or manually annotated readability datasets: MBJP, Buse,
Scalabrino, Dorn, Schnappinger, and JetBrains. We additionally report the
complete 18-feature embedding-only model obtained before causal-LM
features were introduced; these predecessor results are placed
in~\suppref{app:embedding-only-results}. Its reference instance uses
\texttt{jina-embeddings-v2-base-code}, matching the embedding
representation used by \ourModel.

\ourModel{} uses Ridge regression~\cite{hoerl1970ridge} with \(\alpha=200\) to predict
continuous readability scores. Because the six datasets use different rating
scales, their human scores are converted to percentile ranks within each
dataset. During joint training, every dataset receives equal total weight,
preventing larger datasets from dominating \ourModel's fit. Within each evaluation
fold, missing-value imputation, range clipping, and feature standardisation are
estimated from the training partition and applied unchanged to the test
partition. Predictions are bounded to \([0,1]\), with larger values indicating
greater readability.

We use two evaluation protocols: pooled 10-fold cross-validation and
leave-one-dataset-out (LODO) evaluation. In pooled cross-validation, each
dataset is divided into ten folds; in each round, a single Ridge model is
trained on nine folds from every dataset and evaluated on the remaining
folds. The reported correlations are computed exclusively from these
out-of-fold predictions. In LODO evaluation, \ourModel{} is trained on five
complete datasets and evaluated on the sixth.

For each dataset separately, performance is measured using Spearman's
\(\rho\) between predicted and human readability scores. For JetBrains,
the human score is the proportion of reviewers who
judged each snippet readable rather than a derived binary label. We compare \ourModel{} with logical lines of
code (LLOC),
Posnett, Scalabrino, a reconstruction of Dorn, and a reproduction of Mi et
al.'s ConvNetCR model. For the direct-prompting baseline, we apply the
best-performing zero-shot prompt reported by Ou\'edraogo et
al.~\cite{ouedraogo2025llm_method} to DeepSeek-V4-Pro-0813, a
state-of-the-art open model, and GPT-6.1 Sol, a state-of-the-art model.
We query each direct-LLM baseline three times for every sample and use
the mean score in all analyses.
Full specifications appear in~\suppref{app:baselines}.

\subsection{Controlled Readability Interference Experiment}

To examine whether readability models respond in the expected direction
to controlled reductions in readability, we evaluate the frozen \ourModel{}
and the comparison methods on parallel Java and Python interference
datasets constructed from eight well-maintained open-source projects.
The Java corpus is drawn from Guava, Spring Framework, Netty, and Apache
Kafka, while the Python corpus is drawn from Django, Flask, Requests, and
attrs. For each language, we select 25 complete production classes from
each project, resulting in 100 Java classes and 100 Python classes.

We define 13 source-level transformations, organised into seven categories:

\begin{enumerate}
    \item \textbf{Comment transformation:}
    removal of Java line comments, block comments, and Javadoc, and Python
    line comments and docstrings.

    \item \textbf{Identifier transformations:}
    identifier shortening, replacement with length-matched but semantically
    unrelated names, and deterministic minification. Eligible method,
    parameter, and local-variable names follow language-specific binding and
    collision constraints.

    \item \textbf{Expression transformations:}
    hexadecimal rewriting of decimal integer literals and inversion of
    conditions with corresponding branch exchange.

    \item \textbf{Code-injection transformations:}
    insertion of unreachable branches, readable unrelated code, and
    low-readability unrelated code. The latter two are selected
    deterministically from ten template families each.

    \item \textbf{Layout transformation:}
    partial removal of newline-containing whitespace while retaining some
    visual structure, with separate eligibility rules for Python indentation.

    \item \textbf{Data-flow transformations:}
    replacement of integer literals with two short-named local variables whose
    sum preserves the value, and inlining of eligible single-use local
    variables at their sole use site.

    \item \textbf{Control-flow transformation:}
    conversion of eligible \texttt{for} loops into \texttt{while}-based forms,
    covering classic and enhanced loops in Java and synchronous loops through
    explicit iterators in Python.
\end{enumerate}

These transformations reverse readability improvements repeatedly identified
in prior work, including improving names, replacing opaque literals, expressing
conditions positively, simplifying loops, removing unused code, restoring
visual structure, and adding or clarifying documentation~
\cite{buse2009model,scalabrino2018model,dantas2023developers,
sergeyuk2024dataset}. Related perturbation-based studies have likewise used
identifier erosion, misleading replacement, dead-branch injection,
irrelevant-variable injection, and expression rewrites to construct
lower-readability code~\cite{hu2024poor_readability}. All 13 transformations
therefore have a clear expected direction towards lower readability.

The Java and Python datasets use the same 13 transformations, organised
into the same seven categories, with language-specific adaptations where
required. Each transformation
is applied independently to the original class rather than to the output
of another transformation, allowing its effect to be evaluated without
interactions from other interference types.

For each language, the dataset contains 100 original classes and 1,300
independently transformed variants, yielding 1,400 samples per language
and 2,800 samples in total. We take several steps to preserve source consistency wherever possible:
identifier edits follow local bindings, newly introduced names avoid
collisions, and structural transformations update the affected code
within the same source unit. We retain only complete samples that pass
language-specific parsing. Detailed transformation rules, validation
procedures, and eligibility constraints are provided
in~\suppref{app:interference-datasets}.

For a base class \(x_i\), let \(s(x_i)\) denote the readability score
assigned to the original source and let \(s(T_j(x_i))\) denote the score
after applying transformation \(T_j\). We define the paired score difference as
\(\Delta_{ij}=s(T_j(x_i))-s(x_i)\). A negative value indicates that the
transformed version receives a lower
readability score than its matched original. For each transformation, the
response rate is the proportion of changed variants for which
\(\Delta_{ij}<0\).

Only variants whose generated source differs from the original are
included in this calculation; transformations with no applicable target
are excluded. We report response rates separately for Java and Python at
the transformation, category, and overall levels.

%% file: Sections/results.tex
\section{Results}
\label{sec:results}

\ourModel{} contains 11 features selected from the complete candidate
set. We first evaluate its agreement with human ratings across six
readability benchmarks and measure its response to controlled readability
reductions in Java and Python. We then use a human-rated example to contrast
direct LLM judgements with \ourModel{}'s feature-level attribution, before
examining its robustness across
pretrained-model combinations, including the stability of the learned
feature directions. Finally, we evaluate the contribution of its three
feature families and compare it with the independently selected
18-feature embedding-only predecessor.

\subsection{\ourModel{}'s Performance on Readability Benchmarks}
\label{sec:benchmark-results}

We compare \ourModel{} with the readability baselines in~\Cref{tab:benchmark-results}
under pooled 10-fold cross-validation and
leave-one-dataset-out (LODO) evaluation. The reference configuration uses
\nolinkurl{jina-embeddings-v2-base-code} and
\texttt{OpenCoder-1.5B-Base}. Results for all 15 combinations of
embedding and causal language models are reported in~\cref{sec:model-robustness}.

\begin{table*}[t]
\centering
\small
\setlength{\tabcolsep}{3.5pt}
\caption{Performance comparison on readability benchmarks. Each dataset
column reports Spearman's \(\rho\) between predicted and human readability
scores. Average \(\rho\) is the unweighted mean of the six dataset-specific
correlations. Dorn
denotes our reconstruction of the corresponding model, Mi denotes our
ConvNetCR reproduction, and LLOC counts language-aware statement and
declaration nodes. The
shaded block groups the two cloud-LLM baselines: DeepSeek-V4-Pro-0813
and GPT-6.1 Sol.}
\label{tab:benchmark-results}
\begin{tabular}{lccccccc}
\toprule
\textbf{Method / Protocol}
& \textbf{MBJP}
& \textbf{Buse}
& \textbf{Scalabrino}
& \textbf{Dorn}
& \textbf{Schnappinger}
& \textbf{JetBrains}
& \textbf{Average $\rho$} \\
\midrule

Posnett
& 0.500
& \textbf{0.751}
& 0.519
& 0.470
& 0.606
& 0.324
& 0.528 \\

Scalabrino
& 0.332
& 0.690
& 0.602
& 0.586
& 0.414
& 0.342
& 0.494 \\

Dorn
& 0.375
& 0.603
& 0.390
& \textbf{0.618}
& 0.565
& -0.053
& 0.416 \\

Mi
& -0.063
& 0.640
& \textbf{0.626}
& 0.482
& 0.170
& -0.151
& 0.284 \\

LLOC
& -0.590
& -0.136
& -0.359
& -0.229
& -0.571
& -0.279
& -0.361 \\

\rowcolor{cloudllmbg}
DeepSeek
& 0.740
& 0.324
& 0.466
& 0.483
& 0.537
& 0.530
& 0.513 \\

\rowcolor{cloudllmbg}
GPT-6.1
& 0.759
& 0.206
& 0.489
& 0.521
& \textbf{0.646}
& \textbf{0.604}
& 0.537 \\

\midrule

\textbf{\ourModel{} (Pooled CV)}
& 0.851
& 0.671
& 0.549
& 0.561
& 0.625
& 0.531
& 0.631 \\

\textbf{\ourModel{} (LODO)}
& \textbf{0.869}
& 0.674
& 0.560
& 0.552
& 0.633
& 0.540
& \textbf{0.638} \\

\bottomrule
\end{tabular}

\end{table*}

\ourModel{} provides the most consistent performance across the
six datasets, with correlations above 0.53 under both evaluation
protocols. Its average correlations are 0.631 under pooled
cross-validation and 0.638 under LODO, improving over the strongest
baseline average of 0.537 by 0.094 and 0.101, respectively. Individual
baselines remain strongest on Buse, Scalabrino, Dorn, Schnappinger, and
JetBrains, but none
maintains comparable performance across the complete benchmark pool.

The direct-LLM results improve with GPT-6.1 Sol, which obtains the strongest
individual results on Schnappinger and JetBrains. Its average remains 0.537,
however, with correlations ranging from 0.206 on Buse to 0.759 on MBJP.
These results show that strong agreement with one set of readability
judgements does not necessarily transfer to other annotation protocols and
code granularities.

Excluding the target dataset from Ridge training does not reduce overall
performance: the LODO average is slightly higher than the pooled result,
and the lowest dataset-specific correlation increases from 0.531 to
0.540. Although feature selection uses all six development datasets, each LODO
predictor is fitted entirely without samples from its target dataset. Its
performance therefore shows that the selected representation transfers
across datasets without target-dataset fitting or calibration of the predictor.
Dataset-stratified bootstrap confidence intervals are reported
in~\suppref{app:benchmark-uncertainty}. The pooled and LODO lower bounds
are 0.561 and 0.573, respectively, both exceeding the strongest baseline
average point estimate of 0.537.

\Needspace{14\baselineskip}
To quantify sampling uncertainty over this fixed six-dataset benchmark suite,
we use a paired, dataset-stratified bootstrap~\cite{efron1979bootstrap}. Each
replicate resamples the same observations for both methods within each dataset,
recomputes Spearman correlation, and equally averages the six dataset-level
differences to obtain suite-level intervals and tests.

\begin{findingbox}{Answer to RQ1}
\ourModel{} achieves the strongest overall agreement with human readability
judgements under both pooled cross-validation and LODO evaluation, with average
Spearman correlations of 0.631 and 0.638, improving on the strongest baseline
average by 0.094 and 0.101. The corresponding paired 95\% bootstrap intervals
are [0.031, 0.157] and [0.038, 0.165] ($p=0.0032$ and $p=0.0008$), and its
correlations remain above 0.53 on all six datasets.
\end{findingbox}

\subsection{Response to Controlled Readability Degradation}
\label{sec:interference-results}

We next examine whether the frozen \ourModel{} assigns lower readability
scores to the deliberately degraded Java and Python variants.
Response rates by degradation category and language are reported
in~\Cref{tab:interference-results}. A successful response occurs when a
changed variant receives a lower score than its matched original; the
overall rate is calculated over all changed pairs.

\begin{table*}[t]
\centering
\footnotesize
\setlength{\tabcolsep}{1.6pt}
\caption{Response to independently applied readability interferences.
Values denote the percentage of changed variants assigned a lower
readability score than their matched originals. Overall values are
calculated directly over all changed pairs. DeepSeek denotes
DeepSeek-V4-Pro-0813, and GPT-6.1 denotes GPT-6.1 Sol.
The shaded columns group these two cloud-LLM baselines.}
\label{tab:interference-results}
\begin{tabular}{>{\raggedright\arraybackslash}p{0.12\textwidth}
                *{4}{>{\centering\arraybackslash}p{0.105\textwidth}}
                *{2}{>{\columncolor{cloudllmbg}\centering\arraybackslash}p{0.105\textwidth}}
                >{\centering\arraybackslash}p{0.105\textwidth}}
\toprule
\textbf{Category}
& \textbf{Dorn}
& \textbf{Posnett}
& \textbf{Scalabrino}
& \textbf{Mi}
& \textbf{DeepSeek}
& \textbf{GPT-6.1}
& \textbf{\ourModel} \\
\midrule

\multicolumn{8}{l}{\textbf{Java}} \\

Comments
& 90.0\%
& 18.9\%
& 84.1\%
& 46.7\%
& 80.0\%
& 91.1\%
& \textbf{94.4\%} \\

Identifiers
& 55.8\%
& 33.0\%
& 23.7\%
& 28.2\%
& 98.6\%
& \textbf{99.0\%}
& 90.8\% \\

Expressions
& 64.8\%
& 27.3\%
& 66.7\%
& 29.5\%
& 81.8\%
& 75.0\%
& \textbf{92.0\%} \\

Code injection
& 99.6\%
& 40.7\%
& 83.1\%
& 40.3\%
& 99.6\%
& \textbf{100.0\%}
& 98.4\% \\

Layout
& 96.0\%
& 49.0\%
& 96.9\%
& 98.0\%
& \textbf{100.0\%}
& \textbf{100.0\%}
& 99.0\% \\

Data flow
& \textbf{95.6\%}
& 13.3\%
& 40.0\%
& 33.3\%
& 84.4\%
& 86.7\%
& 82.2\% \\

Control flow
& 81.0\%
& 9.5\%
& 85.0\%
& 14.3\%
& 85.7\%
& \textbf{95.2\%}
& 81.0\% \\

\cmidrule(lr){1-8}
\textbf{Overall}
& 79.8\%
& 33.5\%
& 61.4\%
& 41.4\%
& 94.5\%
& \textbf{95.5\%}
& 93.8\% \\

\midrule
\multicolumn{8}{l}{\textbf{Python}} \\

Comments
& 78.3\%
& 40.5\%
& --
& 38.1\%
& 82.1\%
& 88.1\%
& \textbf{92.9\%} \\

Identifiers
& 47.2\%
& 44.3\%
& --
& 29.3\%
& 92.3\%
& \textbf{95.1\%}
& 80.5\% \\

Expressions
& 50.0\%
& 32.6\%
& --
& 10.9\%
& 88.0\%
& 82.6\%
& \textbf{93.5\%} \\

Code injection
& 90.4\%
& 58.9\%
& --
& 29.1\%
& 91.6\%
& 96.1\%
& \textbf{96.8\%} \\

Layout
& 75.6\%
& 56.0\%
& --
& 82.4\%
& 90.1\%
& \textbf{92.3\%}
& 85.7\% \\

Data flow
& 70.4\%
& 14.3\%
& --
& 8.9\%
& 78.6\%
& \textbf{83.9\%}
& 75.0\% \\

Control flow
& 62.5\%
& 20.0\%
& --
& 16.0\%
& 96.0\%
& \textbf{100.0\%}
& 60.0\% \\

\cmidrule(lr){1-8}
\textbf{Overall}
& 68.4\%
& 46.0\%
& --
& 31.8\%
& 89.8\%
& \textbf{92.7\%}
& 87.6\% \\

\bottomrule
\end{tabular}

\vspace{3pt}
\begin{minipage}{0.98\textwidth}
\footnotesize
Scalabrino is omitted from the Python experiment because its released
implementation supports Java only. It produced finite predictions for
97 of the 100 Java groups. The reconstructed Dorn extractor processed
99 of the 100 Python groups. Percentages for these methods are calculated
over their available changed pairs.
\end{minipage}
\end{table*}

\ourModel{} assigns lower readability to 93.8\% of changed Java variants and
87.6\% of changed Python variants. GPT-6.1 Sol reaches 95.5\% in Java and
92.7\% in Python. DeepSeek and \ourModel{} both
exceed 50\% on every transformation in both languages. Dorn remains strongest for
Java data-flow changes, while GPT-6.1 Sol is strongest for Python control-flow
lowering.

\Needspace{14\baselineskip}
For each language, inference is clustered by the 100 original source
classes, preserving the dependence among variants derived from the same
class~\cite{field2007clustered}. We report paired cluster-bootstrap intervals,
and Holm-adjusted tests across the 13 transformations~\cite{holm1979multiple}.

\begin{findingbox}{Answer to RQ2}
\ourModel{} responds in the expected direction to 93.8\% of changed Java
variants and 87.6\% of changed Python variants, within 1.7 and 5.1 percentage
points of the strongest direct-LLM response. It assigns lower readability to a
majority of changed variants across all 13 transformations in both languages.
\end{findingbox}

At the individual-transformation level, both \ourModel{} and DeepSeek
assign lower readability to a majority of changed variants for all 13
transformations in both languages. \ourModel{}'s clearest advantage appears for
integer-literal encoding: it responds correctly to 94.7\% of Java variants
and 100.0\% of Python variants, compared with 60.5\% and 73.3\% for the
strongest baseline in each language. It also leads or ties in both languages
on comment removal and remains highly responsive to both forms of
unrelated-code injection. Complete per-transformation results are reported
in~\suppref{app:interference-details}.

\subsection{\ourModel{}'s Robustness Across Pretrained Models}
\label{sec:model-robustness}

We evaluate \ourModel{} under all 15 pairings of five embedding
models and three causal code language models. We examine whether its
predictive performance and feature-level interpretations remain consistent
across these independently fitted instantiations.

The performance range produced by the five embedding choices for each causal
language model is summarised in~\Cref{tab:model-robustness}.

\begin{table*}[t]
\centering
\small
\caption{Robustness across pretrained-model choices. Each range reports
the minimum and maximum over five embedding-model instantiations for the
corresponding causal language model. Pooled and LODO columns report
average Spearman correlation; Java and Python columns report response
rates over changed variants.}
\label{tab:model-robustness}
\begin{tabular}{lcccc}
\toprule
\textbf{Causal language model}
& \textbf{Pooled $\rho$}
& \textbf{LODO $\rho$}
& \textbf{Java}
& \textbf{Python} \\
\midrule

Qwen2.5-Coder
& 0.613--0.624
& 0.617--0.628
& 94.4--95.8\%
& 81.4--85.8\% \\

DeepSeek-Coder
& 0.623--0.633
& 0.621--0.633
& 94.9--95.4\%
& 79.7--84.7\% \\

OpenCoder
& 0.622--0.631
& 0.627--0.638
& 93.6--95.2\%
& 84.2--88.0\% \\

\bottomrule
\end{tabular}
\end{table*}

Performance varies only modestly across pretrained-model choices. Across
all 15 combinations, the pooled average ranges from 0.613 to 0.633 and
the LODO average from 0.617 to 0.638. Java response rates remain within
93.6--95.8\%, while Python response rates range from 79.7\% to 88.0\%.
The resulting model therefore retains comparable benchmark and
controlled-interference performance when either pretrained component is
changed.

We next test whether changing either pretrained component changes the model's
explanation of readability. We compare the sign and magnitude of every
standardised Ridge coefficient across all 15 causal-LM--embedding-model
pairings in~\Cref{tab:coefficient-stability}. Each pairing is fitted
independently, and no coefficient-sign constraints are imposed.

\begin{table}[t]
\centering
\small
\caption{Stability of standardised Ridge coefficients across all 15
causal-LM--embedding-model pairings. The absolute span is the difference
between the largest and smallest coefficient magnitudes.}
\label{tab:coefficient-stability}
\begin{tabular}{lrr}
\toprule
\textbf{Feature}
& \textbf{Coefficient range}
& \textbf{Absolute span} \\
\midrule

Operator density
& $[-0.0823,\,-0.0777]$
& 0.0047 \\

Literal-tail surprisal
& $[-0.0517,\,-0.0408]$
& 0.0109 \\

Computation-and-control pattern count
& $[-0.0564,\,-0.0497]$
& 0.0066 \\

Identifier-onset surprisal
& $[-0.0324,\,-0.0286]$
& 0.0038 \\

Decision density
& $[-0.0189,\,-0.0149]$
& 0.0040 \\

Expression--literal density
& $[-0.0315,\,-0.0293]$
& 0.0022 \\

Assignment-value surprisal
& $[-0.0273,\,-0.0222]$
& 0.0052 \\

Longest line length
& $[-0.0449,\,-0.0374]$
& 0.0075 \\

Declaration-surprisal variation
& $[-0.0416,\,-0.0234]$
& 0.0183 \\

Short-identifier context dependence
& $[-0.0288,\,-0.0219]$
& 0.0069 \\

Identifier embedding dispersion
& $[-0.0396,\,-0.0309]$
& 0.0087 \\

\bottomrule
\end{tabular}
\end{table}

Every coefficient range lies entirely below zero, giving 100\% direction
agreement across the \(11 \times 15 = 165\) independently fitted
coefficients. The variation is also limited in absolute terms: all spans
are below 0.019 on the standardised coefficient scale, and 10 of the 11
are below 0.011. Changing either pretrained component therefore preserves
the coefficient directions while keeping their values within a limited
absolute range.

This cross-model stability is accompanied by agreement across three
levels of analysis. For every retained feature, the theoretically
expected relationship with readability, the coefficient obtained when
the feature is fitted as the sole predictor, and the coefficient in the
complete multivariate model all have the same direction. The learned
interpretations are therefore consistent both across pretrained-model
choices and between isolated and joint feature analysis.

\begin{findingbox}{Answer to RQ3}
Across all 15 pretrained-model combinations, average benchmark correlation
varies by no more than 0.021, and every combination outperforms the strongest
direct-LLM average under both benchmark protocols. Controlled-transformation
responses remain high across model choices, and no fitted coefficient reverses
direction.
\end{findingbox}

Complete results for all 15 \ourModel{} configurations are reported
in~\suppref{app:model-robustness}.

\subsection{Where Frontier LLM Readability Judgments Fall Short}
\label{sec:model-explainability}

The cloud LLMs approach \ourModel{} on controlled transformations, but these
comparisons favour direct prompting: garbled names, injected code, and
compacted layout make the less-readable variant obvious. Human-rated
benchmarks instead require ranking subtler differences among unmodified
programs, where the LLMs are substantially less reliable. \texttt{Buse/66}
illustrates both this performance gap and the limits of generated explanations.
Rated by 121 participants, the three-statement snippet has a mean human score
of 2.603/5 and lies at the 0.152 readability percentile among the 100 Buse
snippets:

\begin{tcolorbox}[
  colback=black!2,
  colframe=black!25,
  boxrule=0.4pt,
  arc=0.8mm,
  left=1mm,
  right=1mm,
  top=0.8mm,
  bottom=0.8mm,
  before skip=4pt,
  after skip=5pt
]
\begin{minipage}[t]{0.65\linewidth}
\raggedright\footnotesize\ttfamily
mProgramTable.\textcolor{teal!65!black}{changeSelection}(\par
\hspace*{1em}row, \textcolor{orange!80!black}{0},
\textcolor{blue!65!black}{false}, \textcolor{blue!65!black}{false});\par
\textcolor{blue!65!black}{Program} p =
(\textcolor{blue!65!black}{Program}) mProgramTableModel\par
\hspace*{1em}.\textcolor{teal!65!black}{getValueAt}(row,
\textcolor{orange!80!black}{1});\par
\textcolor{blue!65!black}{JPopupMenu} menu = devplugin.Plugin\par
\hspace*{1em}.\textcolor{teal!65!black}{getPluginManager}()\par
\hspace*{1em}.\textcolor{teal!65!black}{createPluginContextMenu}(\par
\hspace*{2em}p, CapturePlugin.\textcolor{teal!65!black}{getInstance}());
\end{minipage}\hfill
\begin{minipage}[t]{0.31\linewidth}
\raggedright\sffamily\scriptsize
\textbf{Readability scores}\par
\begin{tabular}{@{}lr@{}}
Human & 2.603/5 [0.152] \\
DeepSeek & 17/20 [0.874] \\
GPT-6.1 & 18/20 [0.848] \\
\textbf{\ourModel} & \textbf{0.129 [0.152]} \\
\end{tabular}\par
\textcolor{black!60}{Brackets give the Buse rank; for \ourModel{}, the human target.}
\end{minipage}
\end{tcolorbox}

\ourModel{} attributes its low score primarily to operator density ($-0.264$),
literal-tail surprisal ($-0.118$), identifier-onset surprisal ($-0.055$),
assignment-value surprisal ($-0.051$), and longest-line length ($-0.018$).
These penalties identify the dense calls, unusual literals, newly introduced
names, assignment values, and 116-character final line, while their magnitudes
distinguish the relative severity of these problems. In contrast, both
direct LLMs substantially overrate the snippet. DeepSeek calls it generally
clean, while GPT-6.1 identifies the short name \texttt{p}, numeric indices,
unexplained Boolean arguments, and the long final call, yet still assigns
18/20. Across the two responses, the LLMs identify only a subset of the
problems captured by \ourModel{} and fail to reflect either their individual
severity or their combined effect in scores of 17--18/20. The complete
responses are reported in~\suppref{app:buse66-llm-responses}.

Existing feature-based models tie their predictions directly to measured
features, but suffer from other flaws: Dorn and
Scalabrino~\cite{dorn2012model,scalabrino2018model} feature weights flip sign
relative to their domain definitions, while Posnett~\cite{posnett2011model}
relies on just three program-level metrics that yield only coarse explanations.
Crucially, all 11 features
remain directionally consistent—their isolated effects align with their
multivariate fitted weights—ensuring that penalties consistently reflect
readability deficits. Across both examples, \ourModel{} identifies dense calls,
unusual literals, and new names in \texttt{Buse/66}; in
\Cref{fig:motivating-example}, assignment-value and literal-tail surprisal
pinpoint the unusual assignment and encoded string.

\begin{findingbox}{Answer to RQ4}
\ourModel{} explains each prediction through signed contributions from 11 named
source properties. Unlike prior feature-based baselines, every feature in
\ourModel{} has a clear interpretation and is directionally consistent,
ensuring that every feature weight aligns in sign with its expected effect on
readability.
\end{findingbox}

\subsection{Feature Family Ablation}
\label{sec:llm-feature-contribution}

We ablate each of the three feature families from the fixed final
representation without repeating feature selection. After removing a
family, the remaining features are refitted using the same Ridge
hyperparameters, datasets, target transformations, and fold assignments.
We additionally compare \ourModel{} with the independently selected
18-feature embedding-only predecessor.

\begin{table}[H]
\centering
\small
\caption{Feature-family ablation and comparison with the independently
selected embedding-only predecessor. All rows use Jina embeddings;
\ourModel{} configurations use OpenCoder predictability measurements.
Java and Python report response over all changed interference pairs.}
\label{tab:learned-predictability-contribution}
\begin{tabular}{lrrrrr}
\toprule
\textbf{Configuration} & \textbf{Features} & \textbf{Pooled $\rho$}
& \textbf{LODO $\rho$} & \textbf{Java} & \textbf{Python} \\
\midrule
\textbf{\ourModel}
& 11 & \textbf{0.631} & \textbf{0.638} & 93.8\% & \textbf{87.6\%} \\
Without traditional features
& 7 & 0.455 & 0.475 & 88.3\% & 82.1\% \\
Without embedding-derived features
& 9 & 0.587 & 0.596 & \textbf{95.2\%} & 86.7\% \\
Without causal-LM features
& 6 & 0.541 & 0.545 & 73.5\% & 72.9\% \\
\midrule
Embedding-only predecessor
& 18 & 0.551 & 0.548 & 86.8\% & 76.2\% \\
\bottomrule
\end{tabular}
\end{table}

All three families contribute to agreement with human judgements.
Removing the traditional features produces the largest benchmark loss,
reducing the pooled and LODO averages by 0.176 and 0.163. Removing the
causal-LM features reduces them by 0.090 and 0.093 and produces the
largest behavioural loss: the Java and Python response rates fall by 20.2\%
and 14.7\%. Removing the embedding-derived features has a smaller effect on
the human-rated benchmarks, reducing the pooled and LODO averages by 0.045
and 0.042, respectively. Its Python response decreases slightly, whereas its
Java response increases by 1.5\%.

The independently selected predecessor provides a complementary
comparison with the complete representation available before causal-LM
features were introduced. \ourModel{} improves its pooled and LODO
averages by 0.080 and 0.090 and its Java and Python responses by 7.0\% and
11.4\%, while using seven fewer features. The complete
pretrained-model comparison in~\suppref{app:model-robustness} shows
that these gains are not tied to the selected reference pairing.

Full results for the independently selected embedding-only predecessor
are reported in~\suppref{app:embedding-only-results}.

\input{Sections/threats.tex}

%% file: Sections/threats.tex
\subsection{Threats to Validity}
\label{sec:threats}

The six datasets measure readability using different rating scales, human-rating
procedures, and kinds of source units, so their raw scores are not directly
comparable. We convert scores to percentile ranks within each dataset and give
every dataset equal total weight. This preserves the ordering of human judgements
within each dataset while preventing a particular scale or larger dataset from
dominating \ourModel's fit, allowing it to identify readability signals that recur
across independently collected annotation settings.

The human-rated literature is concentrated on Java. This is not a consequence
of selective dataset choice: four of the five public benchmarks are Java-only,
and the fifth includes Java alongside Python and CUDA. To test beyond this
evidence base, we conduct a separate experiment on Python production code,
where \ourModel{} responds in the expected direction to 87.6\% of changed
variants and responds to a majority of variants for every transformation.

The final feature set and its size were selected using all six human-rated
datasets. Pooled cross-validation and LODO consequently evaluate prediction
with this fixed representation, not the transfer of the complete selection
procedure to a future benchmark. We therefore also test the frozen \ourModel{}
across 13 controlled transformations in separately constructed Java and Python
corpora excluded from both feature selection and predictor training.

%% file: Sections/disc.tex
\section{Feature Engineering}
\label{sec:feature-engineering}

We now examine the feature engineering behind \ourModel: how its three feature
families complement one another, how pretrained models provide interpretable
measurements, and what the selection results reveal about the available
supervision.

\subsection{Combining the Three Feature Families}

The results support a view of readability as the joint effect of three
properties: the explicit complexity presented by the source, the semantic
organisation of its program elements, and the extent to which those elements
conform to the context established during reading. The feature family ablation
shows that these properties are complementary rather than interchangeable.
Traditional measurements provide the strongest foundation for agreement with
human ratings, causal-LM measurements contribute both benchmark performance and
sensitivity to controlled degradation, and embedding-derived measurements add
information about organisation that is not fully expressed by either other
family alone. One small exception is the 1.5-point increase in the Java
interference response after removing embedding features. Some controlled
rewrites reduce structural-pattern count or identifier-space dispersion even as
they reduce readability, creating a local mismatch between embedding
compactness and readability.

The change from the embedding-only model to \ourModel{} illustrates how
causal-LM features replace an indirect traditional proxy. Source-text byte
entropy ranked second in the five-embedding consensus, but fell to 39th after
the causal-LM measurements were introduced. Byte entropy captures the overall
irregularity of source text, but cannot show which program element is difficult
to anticipate or whether earlier code helps explain it. The selected
construct-level surprisal features instead localise unexpected information to
literals, identifier onsets, assignment values, declarations, and short
identifiers while quantifying how much preceding context explains it.

\subsection{Pretrained Models as Measurement Instruments}

Direct prompting provides a useful contrast to our use of pretrained models
for feature construction. Newer models improve substantially over earlier
prompt-based results: GPT-6.1 Sol reaches an average correlation of 0.537 and
is strongest on Schnappinger and JetBrains. Its correlations nevertheless
range from 0.206 to 0.759 across the six benchmarks and remain below
\ourModel's average of 0.631. A generated score also does not reveal which
properties of the source shaped the judgement.

The two direct-prompting baselines average 1,083 prompt tokens and 1,245 total
tokens per evaluation. \ourModel{} instead computes its embedding and surprisal
features using compact local models, without remote prompting or response
generation.

Compared with end-to-end neural predictors and earlier feature-based
models, \ourModel's representation follows an information-theoretic view of code
comprehension. Each selected feature describes a recognisable source of
reading difficulty, such as dense operations, fragmented structural
patterns, unexpected identifiers, or values that are difficult to
anticipate from preceding code. For every feature, its expected
relationship with readability, its association when used alone, and its
coefficient in \ourModel's complete fit have the same direction. This
three-way agreement connects each interpretation to both the feature
definition and the empirical evidence.

Among the evaluated readability models, Posnett is the only one that is
both more compact than \ourModel{} and fully consistent between feature meaning
and coefficient direction. Its three features provide useful but broad
summaries, whereas our 11 features identify construct-specific sources of
reading difficulty and achieve substantially stronger agreement with human
ratings.

\subsection{What Feature Selection Reveals About Supervision}

Some readability properties do not appear in \ourModel{}, possibly because
the available human ratings do not capture them consistently. Comments provide
the clearest example: among 12 candidates covering comment presence,
comment--identifier alignment, comment--code relevance, and predictive value,
the highest-ranked is only 44th in the 15-model consensus.

Nevertheless, \ourModel{} responds strongly to comment removal: deleting
comment-only lines raises operator and decision density, while losing preceding
comment context can change the predictability of later program elements. The
retained features therefore capture part of the information supplied by comments
without a dedicated comment feature.

A diagnostic separating comments into six content types still produced no
stable comment-quality feature (see~\suppref{app:comment-diagnostic}). This
may reflect their different functions across datasets, including task
requirements, API documentation, implementation explanations, licence
information, and disabled code.

%% file: Sections/relwork.tex
\section{Related Work}
\label{sec:related-work}

In this work, code readability refers to how easily developers can read
and understand a source-code sample, as reflected in human judgements
\cite{buse2009model,dorn2012model,sergeyuk2024dataset}.

Early automated approaches relied on structural complexity measures such as
Cyclomatic Complexity~\cite{mccabe1976complexity} and Halstead
metrics~\cite{halstead1977complexity}. Although useful for describing program
structure, such metrics provide only indirect proxies for human-perceived
readability and understandability ~\cite{lavazza2023empirical}. However,
Halstead metrics rest on disputed theoretical assumptions, derive some apparent
relationships from their own mathematical definitions, and lack consistent
empirical support
~\cite{lassez1981critical,shen1983software,card1987resolving}. Cyclomatic
Complexity reduces complexity to control-flow decisions without a convincing
link to broader software difficulty, and is strongly confounded with lines of
code, which often performs as well or better
~\cite{shepperd1988cyclomatic,shepperd1994critique}. Similarly, rule-based
heuristics like Cognitive Complexity sum arbitrary syntactic penalties while
ignoring identifier semantics, layout, and visual structure, failing basic
tests of construct validity and formal metric
properties~\cite{LAVAZZA2023111561}. Moreover, readability improvements made
by developers are not always captured by automated static-analysis
tools~\cite{dantas2023developers}.

In parallel, empirical software engineering researchers formulated readability prediction as a
supervised
learning problem. Buse and Weimer~\cite{buse2009model} combined
structural, lexical, syntactic, and layout features to predict human
readability judgements. Posnett et al.~\cite{posnett2011model} showed that
a substantially smaller model based on lines of code, Halstead volume,
and entropy could achieve competitive performance. Scalabrino et
al.~\cite{scalabrino2018model} later incorporated richer textual
properties, including identifier and comment information, while Mi et
al.~\cite{mi2018model1,mi2022model2} explored neural models that reduce
dependence on manually designed features. Across these approaches, models fit
formulas to disputed structural metrics, learn feature effects that conflict
with their stated meanings, or rely on opaque neural representations. They also
remain imperfectly aligned with human
judgements and may
fail to reflect readability improvements perceived by developers
~\cite{pantiuchina2018improving,sergeyuk2024dataset}.

Recent work has investigated language models as an alternative source of
readability signals. Surprisal, defined as the negative log-probability of
an observed item given its preceding context, has a long-standing connection
to expectation-based theories of incremental language-processing difficulty
\cite{levy2008surprisal}. This connection makes language-model probability a
plausible source of readability information, but evidence from complete texts
shows that a single aggregate is not sufficient. For natural-language
sentences, Miaschi et al.~\cite{miaschi2020perplexity} find only a weak
relationship between neural language-model perplexity and readability, with
the two measurements responding to different linguistic phenomena. For
source code, Dillmann et al.~\cite{dillmann2024llm_prob_signal_method}
evaluate whole-class cross-entropy against expert judgements of readability
and related maintainability dimensions. The unadjusted association is
contrary to the expected direction and changes after controlling for
logical lines of code; they conclude that class-level cross-entropy alone
may be insufficient.

A separate line of work uses language models as direct readability
evaluators. Ouédraogo et al.~\cite{ouedraogo2025llm_method} compare direct
readability judgements across models, languages, and prompting strategies,
finding moderate but variable alignment with human ratings.

A complementary line of cognitive research explains understanding in terms of
compact representations: related observations can be expressed through shared
patterns and shorter descriptions~\cite{chater2003simplicity,
feldman2016simplicity}. Research on program comprehension likewise shows that
readers form mental representations at multiple levels of abstraction rather
than retain source details independently~\cite{pennington1987mental_representation,
bidlake2020mental_representations}. This perspective shifts attention from the
amount of source information alone to how readily it can be organised into
coherent higher-level representations.

Our work connects this account of semantic organisation with the established
readability features and language-model evidence reviewed above. Embeddings
measure how source elements organise into related groups, while causal language
models locate information that is unexpected in its preceding context.
\ourModel{} returns to explicit feature-based prediction, but
equips it with pretrained representations used as measurement instruments
rather than direct readability judges.

%% file: Sections/conc.tex
\section{Conclusion}
\label{sec:conclusion}

\ourModel{} shows that code-readability modelling need not trade predictive
strength for interpretability. Using pretrained models as measurement
instruments rather than end-to-end judges, it achieves state-of-the-art human
agreement with stable, interpretable feature directions. This supports
developer feedback, readability-aware LLM post-training, and automated code
improvement.

%% file: Sections/data_availability.tex
\section{Data Availability}

The code, model configurations, recorded results, and dataset retrieval
instructions are available at
\url{https://github.com/WangTairan/BTTF}.
An interactive viewer of the results and feature-level diagnoses is available
at \url{https://wangtairan.github.io/BTTF/}.

%% file: Appendices/features.tex
\section{Feature Extraction and Candidate Feature Set}
\label{app:features}

This appendix specifies the extraction rules and complete candidate
inventories for the three feature families in~\mainpaperref{the main paper's
approach section}{sec:approach}.

\subsection{Feature Extractor}

Each extracted snippet records a normalised string, its source-line
position, and one of the types listed in~\Cref{tab:snippet-taxonomy}.

The snippet representation is language aware. Complete Java and Python
units are processed using their corresponding syntactic representations,
while explicitly declared fragments and C-family code use bounded
language-specific lexical extraction. Extracted strings are
whitespace-normalised and bounded to prevent unusually large constructs
from dominating the embedding computation.

Representative Java examples appear in~\Cref{tab:snippet-taxonomy}.
Because one construct can emit multiple snippets, the extracted strings
shown in the table are illustrative rather than exhaustive.

\begin{table*}[t]
\centering
\small
\caption{Snippet taxonomy and representative extracted strings.}
\label{tab:snippet-taxonomy}
\begin{tabular}{p{0.12\textwidth} p{0.25\textwidth}
                p{0.25\textwidth} p{0.27\textwidth}}
\toprule
\textbf{Type}
& \textbf{Definition}
& \textbf{Source code}
& \textbf{Representative extracted string} \\
\midrule

\texttt{IDENTIFIER}
& Named program entity or reference, such as a variable, field,
parameter, class, method, or constant
& \texttt{return count;}
& \texttt{count} \\

\texttt{DECLARATION}
& Introduction or definition of a program entity, including its name and
available type or parameter information
& \texttt{int count = 0;}
& \texttt{int\_count} \\

\texttt{CALL}
& Function, method, or API invocation
& \texttt{items.size();}
& \texttt{items.size} \\

\texttt{CONTROL\_FLOW}
& Construct that determines or changes execution order
& \texttt{if (count > 0)}
& \texttt{if\_count>0} \\

\texttt{ASSIGNMENT}
& Binding or update of a variable, field, array element, or object
property
& \texttt{sum += value;}
& \texttt{sum+=value} \\

\texttt{ARITHMETIC}
& Numerical computation or arithmetic subexpression
& \texttt{total / count}
& \texttt{total/count} \\

\texttt{COMPARISON}
& Relational, equality, type, or membership predicate
& \texttt{name != null}
& \texttt{name!=null} \\

\texttt{LOGICAL}
& Boolean composition or negation of predicates
& \texttt{ready \&\& !closed}
& \texttt{ready\&\&!closed} \\

\texttt{BITWISE}
& Bit-level operation, shift, or mask
& \texttt{flags \& READ\_ONLY}
& \texttt{flags\&READ\_ONLY} \\

\texttt{LITERAL}
& Constant value appearing directly in source code
& \texttt{return 3;}
& \texttt{literal\_3} \\

\texttt{REGEX}
& Regular-expression component extracted from a string literal used as
a regular expression
& \texttt{Pattern.compile("}\(\backslash\backslash\)\texttt{d+");}
& \texttt{regex\_}\(\backslash\)\texttt{d},
  \texttt{regex\_+} \\

\texttt{IMPORT}
& Standard-library or external dependency referenced by the code
& \texttt{import java.util.List;}
& \texttt{import\_stdlib\_util\_List} \\

\shortstack[l]{\texttt{UNUSED\_}\\\texttt{IMPORT}}
& Imported dependency not referenced by the code
& \texttt{import java.io.File;}
& \texttt{unused\_import\_stdlib\_io\_File} \\

\texttt{COMMENT}
& Natural-language text embedded in source code
& \texttt{// initialize the cache}
& \texttt{comment\_initialize\_the\_cache} \\

\bottomrule
\end{tabular}
\end{table*}

Snippet types and occurrence counts are used to construct traditional
inventory features. Their normalised strings are embedded and used to
construct semantic geometry, clustering, short-identifier context, and
comment-alignment features.

\subsection{Traditional Candidate Features}

The traditional inventory contains 95 features. These measurements are
computed directly from the source text and AST-tagged snippets and do not depend
on an embedding model. The current feature groups are summarised
in~\Cref{tab:base-features}. The counts refer to produced feature columns,
excluding dataset identity, task identifier, and readability-label
columns.

\begin{table*}[t]
\centering
\small
\caption{Traditional candidate feature groups.}
\label{tab:base-features}
\begin{tabular}{@{}p{0.23\textwidth} p{0.62\textwidth} r@{}}
\toprule
\textbf{Feature group}
& \textbf{Representative measurements}
& \textbf{Count} \\
\midrule

Text, lexical, and scale
& Non-empty lines of code, line-length statistics, indentation,
vocabulary size, token count, Halstead volume, comparison and operator
density, and byte entropy
& 13 \\

Visual layout and token distribution
& Indentation transitions, long-line ratios, keyword and punctuation
density, visual-area ratios, normalised token positions, DFT-based
periodicity, per-line token maxima, and aligned visual blocks
& 32 \\

Snippet inventory and geometry
& Snippet count, distinct-snippet ratio, per-line snippet distribution,
snippet-character and snippet-token variation, snippet span,
expression--literal balance, and the available count or ratio measurements
for each snippet type
& 32 \\

Identifier, member-access, and comment structure
& Identifier-length statistics, single-letter and digit-character ratios,
identifier subtoken count, member-access density and chain depth,
comment--identifier word overlap, and line-level identifier coherence
& 12 \\

Compression and redundancy
& Whole-snippet and line-level zlib ratios, five-line-block LZMA ratios,
and cross-line redundancy under zlib and LZMA
& 6 \\

\midrule
\textbf{Total}
& &
\textbf{95} \\

\bottomrule
\end{tabular}
\end{table*}

Each snippet type contributes only the count or ratio measurements defined
for that type. For example, literals are represented through the
expression--literal balance rather than a total-snippet-normalised literal
ratio. The selected \textit{expression--literal density} is the
add-one-smoothed log ratio

\[
\log\!\left(
1+N_{\mathrm{arithmetic}}
+N_{\mathrm{bitwise}}
+N_{\mathrm{comparison}}
+N_{\mathrm{logical}}
\right)
-\log(1+N_{\mathrm{literal}}).
\]

It contrasts compound operations and relations with literal values stated
directly in the source.

\subsection{Embedding-Derived Candidate Features}

The embedding-derived definitions are instantiated independently using
\nolinkurl{nomic-embed-text-v1.5}~\cite{nussbaum2024nomicembedding},
\nolinkurl{Qwen3-Embedding-0.6B}~\cite{zhang2025qwen3embedding},
\nolinkurl{jina-embeddings-v2-base-code}~\cite{gunther2023jinaembedding},
\nolinkurl{snowflake-arctic-embed-m-v2.0}~\cite{yu2024snowembedding}, and
\nolinkurl{voyage-4-nano}~\cite{voyage2026voyage4embedding}.

\paragraph{Snippet views.}

Embedding geometry and clustering are evaluated over four deterministic
views of the snippet sequence. The views separate complementary kinds of
program information without using readability labels.

\begin{table}[t]
\centering
\small
\caption{Snippet views used for embedding-derived features.}
\label{tab:snippet-views}
\begin{tabular}{p{0.27\columnwidth} p{0.64\columnwidth}}
\toprule
\textbf{View} & \textbf{Included snippet types} \\
\midrule

\texttt{all}
& All snippet types \\

\texttt{only\_identifier}
& \texttt{IDENTIFIER} \\

\texttt{semantic\_core}
& \texttt{IDENTIFIER}, \texttt{DECLARATION}, \texttt{CALL}, and
\texttt{LITERAL} \\

\texttt{structural\_core}
& \texttt{CONTROL\_FLOW}, \texttt{ASSIGNMENT}, \texttt{ARITHMETIC},
\texttt{COMPARISON}, \texttt{LOGICAL}, and \texttt{BITWISE} \\

\bottomrule
\end{tabular}
\end{table}

Each distinct type--string pair is embedded once and retains its source
occurrence count as a weight. Transformer outputs are converted to one
vector using attention-mask-aware mean pooling. The vectors are
subsequently normalised before cosine geometry and clustering features
are computed.

For exceptionally large units, feature computation retains at most 512
distinct type--string rows within a view. Selection is deterministic:
higher-frequency rows are retained first, with snippet type and extracted
string used to resolve ties.

\paragraph{Feature inventory.}

The embedding-derived inventory contains 102 features for each embedding
model. The same feature definitions are instantiated independently for
each of the five embedding representations; model-specific matrices are
not concatenated.

The exact number of produced features in each family is reported
in~\Cref{tab:embedding-features}.

\begin{table*}[t]
\centering
\small
\caption{Embedding-derived candidate feature groups.}
\label{tab:embedding-features}
\begin{tabular}{@{}p{0.23\textwidth} p{0.62\textwidth} r@{}}
\toprule
\textbf{Feature group}
& \textbf{Measurements}
& \textbf{Count} \\
\midrule

Targeted semantic alignment
& Comment--identifier cosine alignment, calibrated comment--code
relevance, short-identifier prevalence, and mathematical-versus-application
context
& 9 \\

Snippet-view coverage
& Fraction of source-weighted snippets represented by the
\texttt{only\_identifier}, \texttt{semantic\_core}, and
\texttt{structural\_core} views
& 3 \\

Embedding geometry
& Mean cosine distance to the centroid, pairwise cosine-distance
dispersion, effective rank where applicable, and variance not explained by
the first principal component
& 13 \\

OPTICS
& Unclustered ratio, cluster-size variation, maximum cluster diameter, mean
reachability, and pattern mixing where applicable
& 19 \\

HDBSCAN
& Unclustered ratio, largest-cluster ratio, cluster-size variation, maximum
cluster diameter, and pattern mixing where applicable
& 19 \\

Automatically selected K-means
& Selected cluster count, silhouette score, largest-cluster ratio,
cluster-size variation, and maximum cluster diameter
& 20 \\

Automatically thresholded agglomerative clustering
& Cluster count, largest-cluster ratio, cluster-size variation, maximum
cluster diameter, and pattern mixing where applicable
& 19 \\

\midrule
\textbf{Total}
& &
\textbf{102} \\

\bottomrule
\end{tabular}
\end{table*}

Cluster composition is represented by normalised pattern mixing.

\paragraph{Adaptive clustering.}

The four clustering procedures use deterministic or data-adaptive
configuration rules rather than a fixed-radius DBSCAN sweep. Let \(n\)
denote the number of distinct embedding rows available for a snippet view.

OPTICS~\cite{ankerst1999optics} uses cosine distance, \(\xi=0.05\), and

\[
\operatorname{min\_samples}
=
\min\!\left(
n,\,
\max\!\left(2,\operatorname{round}(\log_2 n)\right)
\right).
\]

HDBSCAN~\cite{campello2013hdbscan} operates on precomputed cosine distances.
Its minimum cluster size and minimum sample count are both set to

\[
\min\!\left(
n,\,
\max\!\left(2,\operatorname{round}(\log_2 n)\right)
\right).
\]

K-means~\cite{macqueen1967kmeans} evaluates

\[
k \in \{2,3,4,6,8,12,16\},
\]

subject to the available number of samples and distinct vectors. The
value maximising cosine silhouette~\cite{rousseeuw1987silhouette} is selected.
Each fit uses ten initializations and a fixed random state.

Agglomerative clustering uses average linkage and cosine distance. Its
distance threshold is set to the median positive pairwise cosine distance
within the current view, allowing the resulting number of clusters to
adapt to the embedding geometry of each source unit.

\paragraph{Targeted semantic measurements.}

Two targeted feature groups address readability cues that cannot be
characterised reliably from surface form alone.

For short identifiers, we first compute

\[
r_{\mathrm{short}}
=
\frac{
N_{\{a,b,c,d,e,i,j,k,m,n,x,y,z\}}
}{
N_{\mathrm{identifier}}
},
\]

where the numerator counts identifiers belonging to the fixed
short-name candidate set. For source units containing at least one such
candidate, code segments containing a candidate are compared with fixed
mathematical and application-code reference centroids. The semantic
context feature is the mean margin

\[
m_{\mathrm{short}}
=
\frac{1}{|S|}
\sum_{s\in S}
\left[
\cos(e_s,\mu_{\mathrm{math}})
-
\cos(e_s,\mu_{\mathrm{application}})
\right],
\]

where \(S\) is the set of candidate-bearing code segments. The feature
records the semantic setting in which those identifiers occur.

Comment-related features are divided into two complementary groups.
First, three features measure alignment between comments and identifiers using the
mean pairwise similarity, maximum pairwise similarity, and similarity
between the weighted comment and identifier centroids. Second, four
features characterise the relevance of comments to code. For each comment, we
compute its maximum cosine similarity to a non-comment semantic snippet.
The resulting similarities are summarised by their weighted mean and
minimum.

To identify potentially irrelevant comments without using readability
labels, each embedding model receives an independently calibrated
similarity threshold. Calibration uses 24 fixed code--comment anchors:
eight Java, eight Python, and eight JavaScript examples. Each anchor
provides one relevant comment and one deterministic mismatched comment
borrowed from another anchor in the same language, producing 48 labelled
calibration comparisons. The threshold maximises balanced accuracy and
is stored with the embedding-model-specific calibration artefact.

The final two comment-relevance features are the source-weighted fraction
of comments below this threshold and their mean positive similarity
deficit. The calibration corpus is fixed independently of the six
readability benchmarks and is not fitted using human readability scores.

\subsection{Causal-LM Predictability Candidate Features}

The extended representation adds 47 canonical feature definitions,
instantiated independently with three frozen causal code language models:
\nolinkurl{Qwen2.5-Coder-0.5B}~\cite{hui2024qwen2models},
\nolinkurl{deepseek-coder-1.3b-base}~\cite{guo2024deepseekmodel}, and
\nolinkurl{OpenCoder-1.5B-Base}~\cite{huang2025opencodermodel}. The models assign teacher-forced
probabilities to the observed source, from which the predictability
measurements are computed.

Let \(p(t_i\mid t_{<i})\) denote the teacher-forced probability of the
observed token \(t_i\). Its surprisal in bits is

\[
s_i=-\log_2 p(t_i\mid t_{<i}).
\]

Because a token can cover a different number of source bytes under
different tokenisations, most aggregate measurements use bits per byte
(BPB). For a source region \(R\),

\[
\operatorname{BPB}(R)=
\frac{\sum_{i\in R}s_i}
     {\text{number of covered UTF-8 source bytes in }R}.
\]

Token losses are aligned back to the original source offsets. When a
token overlaps only part of a measured region, its surprisal is allocated
uniformly across its covered source bytes. Unions of source spans are
used for denominators so that overlapping syntactic roles do not count
the same byte twice within one measurement.

The complete 47-feature inventory is summarised
in~\Cref{tab:causal-lm-features}. Comments are deliberately excluded
from the generic role-difficulty group: the difficulty of predicting
natural-language comment text does not by itself establish whether the
comment helps a reader understand the code. Comments instead contribute
through the two code-prediction-gain measurements described below.

\begin{table*}[t]
\centering
\small
\caption{Causal-LM predictability candidate feature groups.}
\label{tab:causal-lm-features}
\begin{tabular}{@{}p{0.20\textwidth} p{0.66\textwidth} r@{}}
\toprule
\textbf{Feature group}
& \textbf{Measurements}
& \textbf{Count} \\
\midrule

Global difficulty
& Complete-source perplexity, complete-source BPB, and the mean BPB of
the most difficult 20\% of local token blocks
& 3 \\

Typed-unit difficulty
& BPB mean, difficult-tail mean, and occurrence-level standard deviation
for identifiers, declaration headers, call targets, control-flow headers,
assignment right-hand sides, expressions, and literals
& 21 \\

Local organisation
& Block-level BPB standard deviation, tail-to-global ratio, mean and
90th-percentile adjacent-block jumps, longest difficult-block run, and
Gini concentration
& 6 \\

Identifier introduction and reuse
& First-occurrence and reuse BPB, difficult reuse tail, positive reuse
difficulty increase, within-binding variation, short-identifier BPB, and
first-subtoken and continuation-subtoken BPB
& 8 \\

Contextual support
& Longer-context gain for identifiers, short identifiers, call targets,
expressions, assignment right-hand sides, and control-flow headers;
10th-percentile identifier gain; and two measures of the predictive
information supplied by preceding comments
& 9 \\

\midrule
\textbf{Total} & & \textbf{47} \\
\bottomrule
\end{tabular}
\end{table*}

The typed-unit spans are extracted from the source syntax rather than
from tokenizer strings. Declaration headers exclude bodies and variable
initializers; call targets exclude argument expressions; control-flow
headers exclude their controlled bodies; and assignment measurements use
only the initializer or right-hand-side expression. Literal spans exclude
documentation strings. For each role, the mean BPB is weighted by
covered bytes, the tail is the mean of the most difficult 20\% of
occurrences, and the standard deviation measures heterogeneity across
occurrences.

Local-organisation features divide the source trace into consecutive
32-token blocks. They distinguish uniformly difficult code from code in
which surprisal is concentrated in a few regions or changes abruptly.
The difficult-block run uses the within-source 80th percentile and is
normalised by the number of blocks.

Identifier-reuse features operate on confirmed parameter and local-variable
bindings. Identical spellings in different scopes are not merged. The
first occurrence of a binding is separated from its later uses, allowing
the representation to measure whether a name becomes easier to predict
after it has been introduced. First-tokenizer-piece and continuation-piece
measurements allocate surprisal back to the bytes belonging to the same
identifier, reducing dependence on the number of tokenizer pieces.

Contextual-support features compare the same target tokens under two
prefix lengths. Targets are evaluated in 32-token blocks with either 32
or 256 preceding source tokens. For an eligible occurrence,

\[
g(R)=\operatorname{BPB}_{\mathrm{short}}(R)
-\operatorname{BPB}_{\mathrm{long}}(R),
\]

so a positive value means that the additional preceding context makes
the identical target easier to predict. Targets near the beginning of a
source unit, for which the two contexts contain no additional information,
are treated as unavailable rather than assigned a zero gain.

For an eligible preceding comment, the comment measurement keeps the
subsequent code target fixed and compares its BPB with the comment
retained and removed. One feature averages this gain over eligible
comment--code pairs. The second sums the supplied prediction bits and
normalises them by the complete source size, producing a whole-source
density. Source units without an eligible comparison have a missing
pairwise gain; their source-density value is structurally zero because
no measured prediction bits are supplied by comments.

Complete-source scoring uses overlapping 512-token windows with a stride
of 384 tokens while assigning each original token a single loss. Values
requiring a syntactic role or a valid matched comparison are recorded as
missing when that evidence is absent.
Missing-value imputation is performed later within each predictor-training
partition as described in~\mainpaperref{the main paper's evaluation
section}{sec:evaluation}.

%% file: Appendices/feature_sele.tex
\section{Feature Selection Details}
\label{app:selection}
Feature selection consists of candidate eligibility, stability ranking
within each pretrained-model instantiation, and consensus ranking across
independent screens. The embedding-only and extended passes contain 181
and 228 eligible candidates, respectively.

\subsection{Candidate Eligibility}

Several traditional candidates adapt measurements proposed by Dorn and
Scalabrino~\cite{dorn2012model,scalabrino2018model}. We retain them in the
produced feature tables but exclude 16 measurements from screening when
their interpretation depends on snippet presentation, language-specific
punctuation, or an unstable clustering partition:

\begin{itemize}
    \item three cluster-size coefficients of variation: OPTICS over the
    \texttt{only\_identifier} and \texttt{structural\_core} views, and
    automatically selected K-means over the \texttt{semantic\_core} view;
    \item ten absolute vertical-position measurements for identifiers,
    keywords, operators, periods, comments, numeric literals, and snippets;
    and
    \item period density, period-position variation, and DFT-based comma
    periodicity.
\end{itemize}

Only DFT-based comma periodicity would enter the first 11 positions of the
unrestricted extended ranking. It remains excluded because it depends on
the line-wise distribution of a language-specific punctuation mark. The
policy removes three embedding-derived and thirteen traditional candidates,
leaving

\[
(102-3) + (95-13) = 99 + 82 = 181
\]

eligible features in the embedding-only pass. All 47 causal-LM candidates
are eligible, giving

\[
99 + 82 + 47 = 228
\]

features for each extended causal-LM--embedding-model matrix. The 99
embedding-derived columns are recomputed for each embedding model, and the
47 causal-LM columns are recomputed for each causal LM.

\subsection{Stability-based Feature Ranking}

For each screened matrix, we independently rank either 181 or 228
eligible candidates using an L1-regularised logistic
regression~\cite{tibshirani1996lasso}.
Readability scores are binarised independently within every dataset at
the dataset-specific median: samples with scores greater than or equal
to the median are assigned to the higher-readability class. Median ties
remain in that class, and no middle-score samples are discarded. This
produces a common screening task over all 1,100 human-rated samples while
preventing differences in numerical annotation scales from defining the
screening labels.

Missing feature values are imputed using the median estimated from the
corresponding fitting sample, after which the features are standardised.
Samples are weighted inversely by the size of their source dataset so
that each dataset contributes equal total weight. The logistic regression
additionally uses balanced class weights.

The model uses \(C=0.08\), L1 regularisation, the \texttt{liblinear}
solver, a maximum of 5,000 iterations, and random seed 42. To reduce
sensitivity to sampling variation, we perform 200 stability-selection
rounds~\cite{meinshausen2010stability}. In each round, 70\% of the samples
from each of the six datasets are selected uniformly without replacement.
A new imputer, scaler, and logistic regression model are fitted to the
resulting sample.

For each feature \(j\), we compute its selection frequency
\(\pi_j\), defined as the fraction of completed rounds in which its
coefficient satisfies
\[
|\hat{\beta}_{j}| > 10^{-12}.
\]
We additionally record the mean absolute coefficient across the
stability-selection rounds and the absolute coefficient obtained by
fitting the same model to the complete six-dataset screening pool.

Within each embedding-model instantiation, features are ranked
lexicographically in descending order according to
\[
\left(
\pi_j,\;
|\hat{\beta}^{\mathrm{full}}_j|,\;
\frac{1}{200}\sum_{r=1}^{200}
|\hat{\beta}^{(r)}_j|
\right).
\]
Thus, selection frequency is the primary ranking criterion, the
full-pool absolute coefficient is the first tie-breaker, and the mean
absolute coefficient across stability rounds is the second tie-breaker.
The canonical feature name is the final deterministic tie-breaker.

\subsection{Consensus Ranking Across Pretrained Models}

Each screened matrix produces an independent ranking of the same eligible
feature definitions: 181 in the embedding-only pass and 228 in the
extended pass. The feature matrices are not concatenated. Instead,
features with the same canonical definition are matched by name. The
embedding-only consensus aggregates five embedding-model rankings. The
extended consensus aggregates 15 rankings obtained from the Cartesian
product of three causal LMs and five embedding models.

For feature \(j\) and screened model pairing \(m\), let \(\pi_{jm}\) denote its
stability-selection frequency and let
\(\hat{\beta}^{\mathrm{full}}_{jm}\) denote its coefficient in the
complete-pool logistic-regression fit. We define the feature as active
under model \(m\) when

\[
a_{jm}
=
\mathbb{I}
\left[
\pi_{jm}>0
\;\lor\;
\left|\hat{\beta}^{\mathrm{full}}_{jm}\right|>10^{-12}
\right].
\]

Its cross-model support is then

\[
A_j=\sum_{m=1}^{M}a_{jm},
\]

where \(M=5\) for the embedding-only pass and \(M=15\) for the extended
pass. Thus, cross-model support does not indicate membership in a separately
chosen per-model Top-\(k\) set. Instead, it indicates the number of
independently screened model pairings under which a feature receives
nonzero support.

For every candidate feature, we additionally compute its mean selection
frequency, mean absolute complete-fit coefficient, mean reciprocal rank,
and best individual-model rank. The final consensus ranking orders
features lexicographically according to:

\begin{enumerate}
    \item cross-model support \(A_j\) (descending);
    \item mean selection frequency across the \(M\) screened matrices
    (descending);
    \item mean absolute coefficient from the \(M\) complete-pool fits
    (descending);
    \item mean reciprocal rank across screened matrices (descending);
    \item the best rank achieved under any individual screened matrix
    (ascending); and
    \item canonical feature name (ascending).
\end{enumerate}

The support count first distinguishes features that are consistently
active across pretrained-model pairings. Because many candidates receive
nonzero support from all screened matrices, the mean selection frequency
provides the principal discrimination within the largest support group.
The remaining criteria resolve further ties using coefficient magnitude
and consistency across the per-model rankings.

The support distribution for the extended screening pool is reported
in~\Cref{tab:active-count}. All feature families are matched by canonical
definition across the 15 rankings; causal-LM values are separately measured
by each causal LM.

\begin{table}[H]
\centering
\small
\caption{Distribution of the 228 eligible extended candidates by support
across the 15 causal-LM--embedding-model screens.}
\label{tab:active-count}
\begin{tabular*}{0.72\textwidth}{@{\extracolsep{\fill}}rr@{\hspace{3em}}rr@{}}
\toprule
\textbf{Active screens}
& \textbf{Candidates}
& \textbf{Active screens}
& \textbf{Candidates} \\
\midrule
15 & 133 & 7 & 4 \\
14 & 6   & 6 & 7 \\
13 & 10  & 5 & 4 \\
12 & 15  & 4 & 1 \\
11 & 5   & 3 & 5 \\
10 & 10  & 2 & 2 \\
9  & 3   & 1 & 2 \\
8  & 5   & 0 & 16 \\
\midrule
\multicolumn{3}{r}{\textbf{Total}} & \textbf{228} \\
\bottomrule
\end{tabular*}
\end{table}

In the embedding-only pass, 120 of 181 candidates receive nonzero
support from all five embedding representations. In the extended pass,
133 of 228 candidates receive support in all 15 screens. Consequently,
cross-model support alone is
not sufficient to determine either ordering; selection frequency and the
subsequent tie-breaking criteria remain important within the largest
support group.

\subsection{Selection Outcomes and Feature Budget}

\paragraph{Embedding-only pass.}

The first pass was completed before causal-LM features were introduced.
We reviewed the leading consensus-ranked candidates for interpretability,
portability, redundancy, sensitivity to extreme observations, and fitted
direction. Pairwise redundancy was defined by an absolute Pearson or
Spearman correlation above 0.9. The resulting independently selected
model contains 18 features; all occur within the first 24 consensus
positions and receive nonzero support from all five embedding models.
Its cross-embedding performance is reported
in~\Cref{app:embedding-only-results}.

\paragraph{Extended pass.}

\begin{figure}[H]
\centering
\includegraphics[width=\textwidth]
{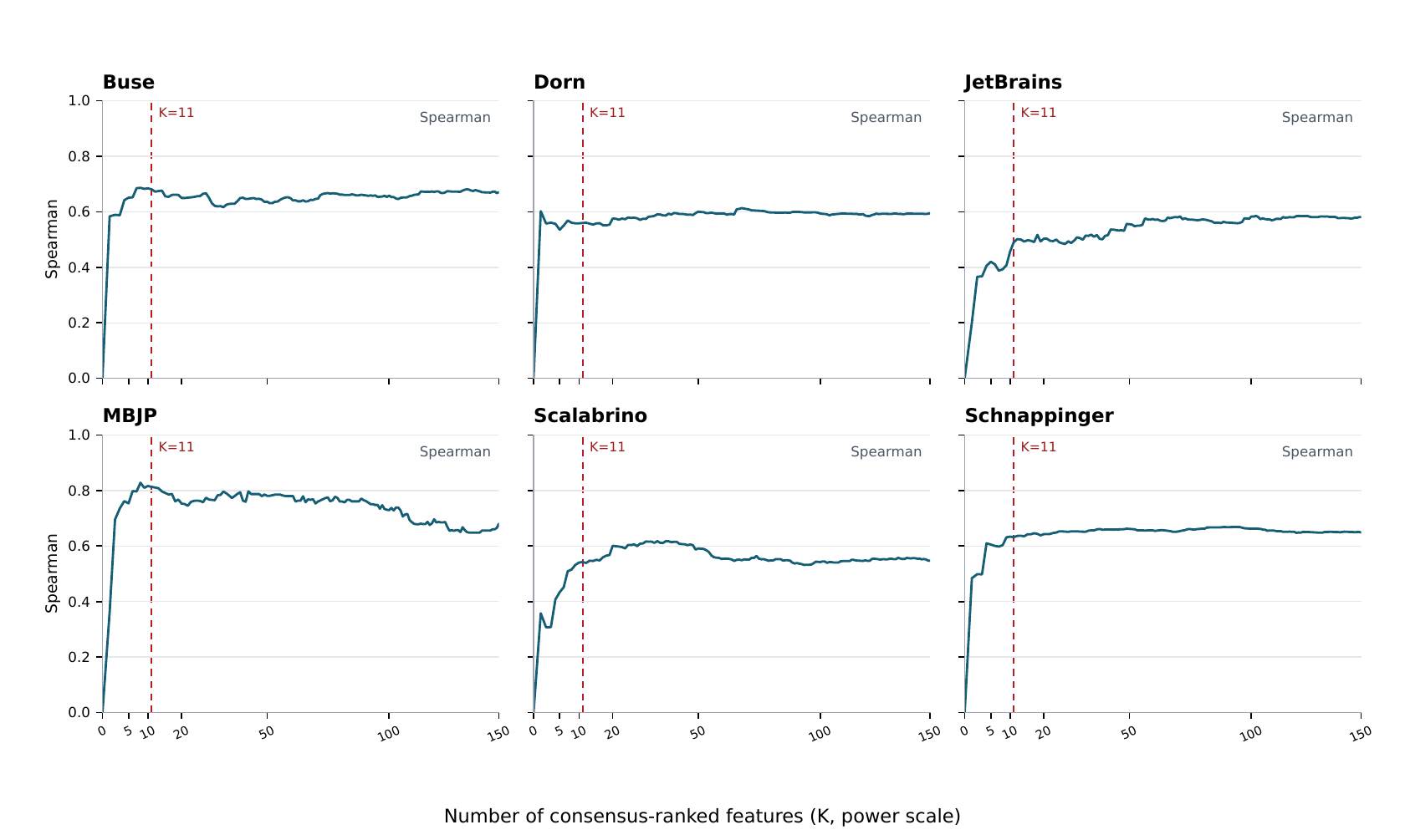}
\caption{Pooled 10-fold cross-validation performance for prefixes of the
extended consensus ranking. The ranking is fixed before this diagnostic
evaluation. At each prefix size, each curve averages the three
causal-LM-specific Ridge results. The dashed line marks the final feature
budget (\(K=11\)).}
\label{fig:feature-budget}
\end{figure}

After adding the 47 causal-LM candidates, we rerun all 200 stability
rounds for each of the 15 causal-LM--embedding-model pairings and
construct a new 228-feature consensus ranking. Successive prefixes of this
fixed extended ranking are evaluated in~\Cref{fig:feature-budget} using pooled
10-fold cross-validation on the six human-rated development datasets. At
each prefix size, the plotted value for a dataset is the mean of three
separately fitted predictors, one for each causal LM, using the Nomic
embedding instantiation; predictions are not ensembled.

Performance rises rapidly in the leading portion of the ranking and then
enters a less stable region in which additional features produce smaller,
dataset-dependent changes. We retain the first 11 positions as the final
operating point. The final representation is therefore the literal first
11 positions of the extended consensus ranking. A redundancy pass at the
predefined absolute Pearson-or-Spearman
threshold of 0.9 rejects no feature in this prefix.

The selected prefix contains four traditional, two embedding-derived,
and five causal-LM predictability features. Every instantiation uses the
same 11 canonical definitions; only the pretrained-model-dependent
measurements change. The complete definitions and interpretations appear
in~\mainpaperref{the main paper's approach section}{sec:approach}
and~\Cref{app:features}.

\subsection{Content-Aware Comment Diagnostic}
\label{app:comment-diagnostic}

We additionally examine whether separating comments by content yields a
more useful comment-predictability signal. Comments are grouped
deterministically as descriptive prose, code-like content,
licence/provenance metadata, visual separators, mixed content, or
unclassified content. For each group, we add its mean comment BPB as a
candidate and repeat the same 15-model stability-selection procedure,
expanding the candidate pool from 228 to 234 features.

\begin{table}[H]
\centering
\small
\caption{Consensus ranks of the content-aware comment-predictability
candidates. Observed samples contain at least one comment in the
corresponding group.}
\label{tab:comment-content-diagnostic}
\begin{tabular}{lrr}
\toprule
\textbf{Comment content group}
& \textbf{Observed samples}
& \textbf{Consensus rank} \\
\midrule
Mixed content              & 159 & 23  \\
Descriptive prose           & 486 & 43  \\
Unclassified content        & 177 & 66  \\
Code-like content           & 80  & 105 \\
Visual separators           & 16  & 164 \\
Licence/provenance metadata & 193 & 216 \\
\bottomrule
\end{tabular}
\end{table}

None of the six candidates enters the first 11 positions, and the
membership of the selected feature set remains unchanged. Separating
comments by content therefore does not produce a dedicated
comment-predictability feature with sufficient independent support to
enter the final representation.

%% file: Appendices/reada_dataset.tex
\section{Human-Rated Dataset Details}
\label{app:datasets}

The six benchmark targets are derived from their released human ratings
as follows.

\subsection{Published Datasets}

For Buse, Scalabrino, and Dorn, we use the arithmetic mean of the
individual human ratings released with each
dataset~\cite{buse2009model,scalabrino2018model,dorn2012model}. Buse
contains 121 ratings for every snippet, while Scalabrino contains nine
ratings for every method. The number of ratings in Dorn varies across
snippets, ranging from 161 to 267.

Schnappinger provides an expectation-maximisation posterior over four
ordered readability classes~\cite{schnappinger2020dataset}. Given the
posterior vector
$(p_0,p_1,p_2,p_3)$, where the classes range from 4 (most readable) to
1 (least readable), we compute the continuous target as

\[
y = 4p_0 + 3p_1 + 2p_2 + p_3.
\]

For JetBrains~\cite{sergeyuk2024dataset}, we use the fraction of votes that
classify a snippet as readable:

\[
y =
\frac{N_{\mathrm{readable}}}
     {N_{\mathrm{readable}} + N_{\mathrm{unreadable}}}.
\]

The number of votes per JetBrains sample ranges from 11 to 31, with a
median of 20. Although the dataset also supports a binary majority-vote
label, we retain the vote fraction as a continuous target for the final
regression and evaluation tasks so that the degree of annotator agreement
is preserved. During feature screening, this continuous target is
binarised at its dataset-specific median in the same manner as the other
five datasets; the released majority-vote label is not used. All targets
are oriented such that larger values indicate greater readability.

\subsection{MBJP Development Dataset}
\label{app:datasets-mbjp}

MBJP is a small development dataset constructed specifically for this
study. We select programming tasks from the Java portion of
MBXP~\cite{athiwaratkun2023multilingual} and use
\texttt{gpt-4.1-nano-2025-04-14} to generate candidate Java solutions.
The solutions are generated under two prompting conditions intended to
produce contrasting readability characteristics while preserving
functional correctness.

The high-readability condition asks the model to act as a proficient Java
developer and produce a clear implementation:

\begin{tcolorbox}[
    title={High-readability generation prompt},
    breakable,
    colback=white,
    colframe=black!40,
    boxrule=0.5pt,
    arc=1mm,
    left=2mm,
    right=2mm,
    top=1.5mm,
    bottom=1.5mm
]
\small\ttfamily
You are a proficient Java developer. Generate highly readable Java code
that correctly satisfies the following task. Return only the complete
Java code.

\medskip
Task:

{\normalfont\{}MBJP task description{\normalfont\}}
\end{tcolorbox}

The low-readability condition asks the model to produce a correct but
deliberately difficult-to-read implementation. The prompt identifies
confusing control flow, ambiguous identifiers, and inappropriate comments
as representative sources of low readability:

\begin{tcolorbox}[
    title={Low-readability generation prompt},
    breakable,
    colback=white,
    colframe=black!40,
    boxrule=0.5pt,
    arc=1mm,
    left=2mm,
    right=2mm,
    top=1.5mm,
    bottom=1.5mm
]
\small\ttfamily
You are a Java learner. Generate low-readability Java code that correctly
satisfies the following task. The code must remain functionally correct.

\medskip
Low-readability code may contains:\\
- confusing or unnecessarily control flow;\\
- strange or meaningless variable names;\\
- other characteristics that make the code harder to read.

\medskip
Task:

{\normalfont\{}MBJP task description{\normalfont\}}
\end{tcolorbox}

Each generated program is checked using the task-specific validation
procedure provided by MBJP. Programs that fail validation are discarded
rather than repaired or assigned a readability score. After this
filtering step, the final development dataset contains 17 functionally
correct Java solutions: ten generated under the high-readability
condition and seven under the low-readability condition.

The prompting condition is used only to introduce variation in the
generated code and is not treated as the readability ground truth.
Instead, each retained solution is independently evaluated by two human
annotators on a five-point readability scale. The mean of the two ratings
is used as the continuous target. The annotators achieve a Spearman
correlation of \(0.829\) and an ICC(2,k) of \(0.899\), indicating strong
agreement~\cite{koo2016icc}.

%% file: Appendices/baseline.tex
\section{Baseline Implementations}
\label{app:baselines}

We compare \ourModel{} against three conventional readability baselines
(LLOC, Posnett, and Scalabrino), two reconstructed learned baselines
(Dorn and Mi), and three direct LLM-based baselines. Fixed
published models are evaluated using their released implementations or
coefficients, whereas the reconstructed baselines are trained according
to the protocols described below.

\subsection{Conventional and Reconstructed Baselines}

Logical lines of code (LLOC) is the number of language-aware statement and declaration nodes in
the source AST; comments and blank lines are not counted. Java and Python
use their respective grammars, while C, C++, and CUDA share the C++ grammar.
For Posnett et al.~\cite{posnett2011model},
we use the published model without retraining:

\[
z = 8.87 - 0.033V + 0.40L - 1.5H,
\qquad
P(\mathrm{readable}) = \frac{1}{1+e^{-z}},
\]

where \(V\) is Halstead volume, \(L\) is the number of source lines, and
\(H\) is source-code byte entropy. We evaluate the comprehensive model of
Scalabrino et al.~\cite{scalabrino2018model} using the authors' released
Java archive and pretrained classifier. Method-level Java snippets are
placed in a minimal enclosing class when required by the released tool,
and samples that cannot be processed are omitted rather than assigned
fallback predictions.

Because no pretrained Dorn model was released, we reconstruct the model
of Dorn and Weimer~\cite{dorn2012model} using the public Dorn feature
data and the released implementations of its visual metrics. Metrics
defined for fewer than 80\% of the training samples are excluded, seven
features are selected through a forward wrapper evaluated with stratified
10-fold accuracy, and a logistic-regression model is fitted on all 360
Dorn samples. The readable-class probability is used as the continuous
readability score.

We also reconstruct the character-level ConvNetCR branch described by
Mi et al.~\cite{mi2018model1}. This reproduction follows the published
convolutional architecture but is not presented as the complete DeepCRM
model, whose full implementation and trained weights were not released.
The model contains three line-spanning convolution banks with 100 feature
maps per bank, followed by ReLU activation, global max pooling, dropout,
and a two-class output layer. It is trained on the highest and lowest
readability quartiles of Buse, the Java subset of Dorn, and Scalabrino,
with the middle half of each score distribution excluded. The resulting
readable-class probability is used as its continuous score.

\subsection{Direct LLM Baselines}

Ouédraogo et al.~\cite{ouedraogo2025llm_method} systematically compare
zero-shot, few-shot, chain-of-thought, and tree-of-thought prompting for
code readability assessment. In their study, zero-shot prompting obtains
the lowest mean absolute error with respect to human readability scores.
We therefore adopt their developer-guided zero-shot formulation as our
direct LLM baseline and instantiate it with DeepSeek-V4-Pro-0813, GPT-6
Sol, and GPT-6.1 Sol. The main paper reports DeepSeek-V4-Pro-0813 and
GPT-6.1 Sol; the complete GPT-6 Sol results are retained in the supplementary
material.

\begin{tcolorbox}[
    title={Direct LLM Readability Prompt},
    breakable,
    colback=white,
    colframe=black!40,
    boxrule=0.5pt,
    arc=1mm,
    left=2mm,
    right=2mm,
    top=1.5mm,
    bottom=1.5mm
]
\small\ttfamily
You are a software engineer evaluating code readability.

\medskip
Evaluate the following code on a scale from 0 (unreadable) to 20
(highly readable), explicitly using these criteria:

\medskip
1. Code Structure - logical organisation vs tangled code\\
2. Nesting - flat vs deeply nested\\
3. Clarity of intent - clear purpose vs ambiguous\\
4. Code length - concise vs unnecessarily long\\
5. Action granularity - one action per line vs multiple\\
6. Reading flow - natural vs disorganised

\medskip
Return ONLY JSON:

{\normalfont\{}
  "score": "0-20",\\
  "reasoning": "concise explanation (max 200 words)"
{\normalfont\}}

\medskip
Code:

{\normalfont\{}code{\normalfont\}}
\end{tcolorbox}

The same prompt is applied to every sample without demonstrations or
access to the corresponding human readability labels. Each model scores every
sample three times, and we report the mean. GPT-6 Sol and GPT-6.1 Sol are evaluated through
the OpenAI API using the model identifiers \texttt{gpt-6-sol} and
\texttt{gpt-6.1-sol}. GPT-6 uses no additional reasoning effort, while GPT-6.1
uses low reasoning effort, the minimum level supported by that model. None of
the models receives tools or executes an agentic workflow; each source unit is
judged independently from the fixed prompt. The experiments were run on
October 2--3, 2026, and the returned
model identifiers, scores, explanations, token usage, and request parameters
are retained with the per-sample results.

\subsection{Direct-LLM Responses for the \texttt{Buse/66} Case}
\label{app:buse66-llm-responses}

\mainpaperref{Section~4.4}{sec:model-explainability} examines a snippet that receives a mean
human rating of 2.603/5 but substantially higher scores from all three direct
LLM baselines. Representative returned explanations are reproduced below; the
headings give the score accompanying each displayed response.

\paragraph{DeepSeek-V4-Pro-0813 (16/20).}
\begin{quote}
\small
The code is generally clean and readable, with a logical structure and clear
intent. It performs three distinct actions: selecting a row, retrieving a
program object, and creating a context menu. Each action is on its own line,
which supports good action granularity. Nesting is flat, and the flow is
natural. However, the readability is slightly reduced by the long method call
to create a context menu, which could be more concise or broken down.
Additionally, the use of \texttt{getValueAt(row, 1)} assumes knowledge of the
table model's column indices, which could be made clearer with a named
constant. Overall, the code is straightforward but could benefit from minor
refactoring for improved clarity.
\end{quote}

\paragraph{GPT-6 Sol (18/20).}
\begin{quote}
\small
The three statements follow a logical sequence: select a row, retrieve its
Program, then create a context menu. Nesting is absent, each line performs one
action, and the code is concise. Intent and reading flow are clear, though the
long chained call and numeric column index slightly reduce readability.
\end{quote}

\paragraph{GPT-6.1 Sol (18/20).}
\begin{quote}
\small
Code Structure: Three logically ordered statements, clearly separated.
Nesting: Completely flat, with no nested control flow. Clarity of intent:
Selecting a row, retrieving its Program, and creating a context menu are
apparent, though the variable name p, numeric column indices, and boolean
arguments lack descriptive context. Code length: Concise, with no unnecessary
statements. Action granularity: Each line performs one main action, although
the final statement chains several calls. Reading flow: Natural progression
from selection to data retrieval to menu creation; the long final line
slightly reduces readability.
\end{quote}

%% file: Appendices/degr_dataset.tex
\section{Controlled Readability Interference Datasets}
\label{app:interference-datasets}

\subsection{Construction and Scope}

We construct two controlled readability-interference datasets, one for
Java and one for Python. Each dataset contains 100 production-code
classes selected evenly from four open-source projects. The Java dataset
contains 25 classes from Apache Kafka, Google Guava, Netty, and Spring
Framework. The Python dataset contains 25 classes from Django, Flask,
Requests, and attrs.

For every source class, we independently apply 13 readability
interferences. Each transformed variant is generated directly from the
original source rather than from a previously transformed version.
Consequently, the datasets do not define cumulative degradation levels,
and the response to each interference can be evaluated against the same
original source.

Each language-specific dataset contains 100 original classes and
\(100 \times 13 = 1{,}300\) interference variants, resulting in 1,400
records per language and 2,800 records in total. Some transformations are
not applicable to every class. For example, integer-literal encoding
cannot change a class containing no eligible decimal integer literal.
Such unchanged variants remain in the dataset for completeness. The
evaluation includes only source pairs that differ, yielding 896 changed
Java variants and 920 changed Python variants.

\subsection{Transformation Details}

The 13 interferences cover seven readability dimensions: comments,
identifiers, expressions, code injection, layout, data flow, and control
flow. The transformations and the number of source classes for which each
produced a source-level change are summarised in~\Cref{tab:constructed-interferences}.

\begin{table*}[t]
\centering
\footnotesize
\setlength{\tabcolsep}{2.5pt}
\caption{Independently applied readability interferences. ``Changed''
denotes the number of transformed classes whose source differs from the
corresponding original.}
\label{tab:constructed-interferences}
\begin{tabular}{p{0.12\textwidth} p{0.21\textwidth}
                p{0.43\textwidth} rr}
\toprule
\textbf{Category}
& \textbf{Interference}
& \textbf{Definition}
& \textbf{Java}
& \textbf{Python} \\
\midrule

Comments
& Comment removal
& Removes comments and documentation, including Javadoc in Java and
docstrings in Python.
& 90 & 84 \\

\midrule

Identifiers
& Identifier shortening
& Deterministically shortens eligible method, parameter, and local-variable
names while avoiding naming conflicts.
& 98 & 95 \\

Identifiers
& Misleading identifiers
& Replaces eligible names with length-matched, meaningful but contextually
unrelated identifiers.
& 98 & 96 \\

Identifiers
& Garbled identifiers
& Replaces eligible names with deterministic ProGuard-style or
minifier-style short identifiers.
& 98 & 96 \\

\midrule

Expressions
& Integer-literal encoding
& Rewrites eligible decimal integer literals using equivalent hexadecimal
notation.
& 38 & 30 \\

Expressions
& Condition inversion
& Negates eligible conditions and exchanges the corresponding branches.
& 50 & 62 \\

\midrule

Code injection
& Dead-branch injection
& Inserts deterministic runtime-dead branches into eligible callable
bodies.
& 86 & 95 \\

Code injection
& Readable unrelated-code injection
& Inserts one of ten readable but functionally unrelated code templates
into an eligible callable.
& 86 & 95 \\

Code injection
& Low-readability unrelated-code injection
& Inserts one of ten functionally unrelated templates written with
deliberately poor readability.
& 86 & 95 \\

\midrule

Layout
& Partial layout compaction
& Deterministically removes approximately 40\% of eligible line-break
groups while retaining part of the original visual structure.
& 100 & 91 \\

\midrule

Data flow
& Numeric-intermediate introduction
& Replaces eligible integer literals with the sum of two newly introduced
short-named local variables whose values reconstruct the original literal.
& 28 & 28 \\

Data flow
& Intermediate-variable inlining
& Removes eligible single-use local variables and substitutes their
parenthesised initializer expressions at their sole use sites.
& 17 & 28 \\

\midrule

Control flow
& Loop lowering
& Rewrites eligible \texttt{for} loops into behaviourally corresponding
\texttt{while}-based forms.
& 21 & 25 \\

\midrule
\multicolumn{3}{r}{\textbf{Total changed variants}}
& \textbf{896}
& \textbf{920} \\

\bottomrule
\end{tabular}
\end{table*}

The two unrelated-code injection conditions use separate sets of ten
templates. Readable templates use conventional control flow, formatting,
and descriptive identifiers, whereas low-readability templates use
compact identifiers, opaque expressions, and deliberately less transparent
control flow. Both transformations insert at most one template into a
class, allowing their effects to be evaluated under comparable target
cardinality.

Identifier transformations operate on eligible method, parameter, and
local-variable bindings and update recognised references consistently.
Constructors and language-required names are preserved. Candidate names
that would conflict with existing bindings or language keywords are
rejected. Misleading identifiers are selected primarily by length
proximity to the original name, while garbled identifiers follow a
deterministic shortest-available-name strategy.

Integer-literal encoding rewrites eligible decimal integer literals using
lowercase hexadecimal notation. It does not recursively transform literals
that are already hexadecimal. Numeric-intermediate introduction performs
the complementary data-flow modification: an eligible integer literal is
decomposed into two values stored in newly introduced short-named
variables, and the original literal is replaced by their sum.

Intermediate-variable inlining targets eligible local variables that have
an initializer, are not subsequently reassigned, and have exactly one
supported use. The declaration is removed, and the use is replaced by a
parenthesised copy of the initializer. Language-specific exclusions are
applied where the transformation cannot be performed reliably.

Condition, data-flow, and control-flow transformations are applied only
when the corresponding syntactic structures satisfy their safety
constraints. Java loop lowering covers eligible classic and enhanced
\texttt{for} loops. Python loop lowering covers eligible synchronous
\texttt{for} loops using explicit iterator-based control flow. Therefore,
the number of changed variants differs across transformations and
languages.

\subsection{Dataset Records and Validation}

Both datasets are generated deterministically using the fixed seed
\texttt{20260823}. Every original and generated Java source is decoded as
UTF-8 and parsed with Tree-sitter Java without syntax errors. Every
original and generated Python source is valid UTF-8 and must parse
successfully using the standard-library \texttt{ast.parse}
implementation. A validation failure terminates dataset construction
rather than retaining an invalid variant.

For each record, we store its source project and path, base-sample
identifier, variant identifier, parent identifier, interference category,
interference name, transformation statistics, line count, original content
hash, and transformed content hash. Repository revisions are recorded
where available. The hashes identify unchanged variants without requiring
manual source comparison and support verification of deterministic
regeneration.

The transformations operate locally on complete source classes. Binding-aware
identifier updates, collision checks for introduced names, and
structure-specific rewrite rules preserve source consistency, while parsing
ensures class-level syntactic validity.

\subsection{Fine-Grained Response Results}
\label{app:interference-details}

Response rates for all 13 individual interference types in Java and Python are
reported in~\Cref{tab:interference-details-java,tab:interference-details-python}.

\begin{table*}[t]
\centering
\scriptsize
\setlength{\tabcolsep}{1.8pt}
\caption{Fine-grained response to independently applied readability
interferences in Java. DeepSeek denotes DeepSeek-V4-Pro-0813; GPT-6 and
GPT-6.1 denote GPT-6 Sol and GPT-6.1 Sol.}
\label{tab:interference-details-java}
\begin{tabular*}{\textwidth}{@{\extracolsep{\fill}}>{\raggedright\arraybackslash}p{0.27\textwidth}cccccccc@{}}
\toprule
\textbf{Interference}
& \textbf{Dorn}
& \textbf{Posnett}
& \shortstack{\textbf{Scala-}\\\textbf{brino}}
& \textbf{Mi}
& \textbf{DeepSeek}
& \textbf{GPT-6}
& \textbf{GPT-6.1}
& \textbf{\ourModel} \\
\midrule

Comment removal
& 90.0\%
& 18.9\%
& 84.1\%
& 46.7\%
& 80.0\%
& 91.1\%
& 91.1\%
& \textbf{94.4\%} \\

\addlinespace
Identifier shortening
& 61.2\%
& 36.7\%
& 16.5\%
& 22.4\%
& 95.9\%
& 95.9\%
& \textbf{96.9\%}
& 88.8\% \\

Misleading identifier substitution
& 42.9\%
& 43.9\%
& 54.6\%
& 38.8\%
& \textbf{100.0\%}
& 99.0\%
& \textbf{100.0\%}
& 96.9\% \\

Garbled identifier substitution
& 63.3\%
& 18.4\%
& 0.0\%
& 23.5\%
& \textbf{100.0\%}
& \textbf{100.0\%}
& \textbf{100.0\%}
& 86.7\% \\

\addlinespace
Integer-literal encoding
& 18.4\%
& 26.3\%
& 42.1\%
& 31.6\%
& 60.5\%
& 31.6\%
& 42.1\%
& \textbf{94.7\%} \\

Condition inversion and branch exchange
& \textbf{100.0\%}
& 28.0\%
& 85.7\%
& 28.0\%
& 98.0\%
& 98.0\%
& \textbf{100.0\%}
& 90.0\% \\

\addlinespace
Dead-branch injection
& \textbf{100.0\%}
& 40.7\%
& 97.6\%
& 58.1\%
& \textbf{100.0\%}
& \textbf{100.0\%}
& \textbf{100.0\%}
& 95.3\% \\

Readable unrelated-code injection
& 98.8\%
& 40.7\%
& 84.7\%
& 9.3\%
& \textbf{100.0\%}
& \textbf{100.0\%}
& \textbf{100.0\%}
& \textbf{100.0\%} \\

Low-readability unrelated-code injection
& \textbf{100.0\%}
& 40.7\%
& 67.1\%
& 53.5\%
& 98.8\%
& \textbf{100.0\%}
& \textbf{100.0\%}
& \textbf{100.0\%} \\

\addlinespace
Partial layout compaction
& 96.0\%
& 49.0\%
& 96.9\%
& 98.0\%
& \textbf{100.0\%}
& \textbf{100.0\%}
& \textbf{100.0\%}
& 99.0\% \\

\addlinespace
Numeric-intermediate introduction
& \textbf{100.0\%}
& 21.4\%
& 46.4\%
& 35.7\%
& 92.9\%
& 92.9\%
& \textbf{100.0\%}
& 89.3\% \\

Intermediate-variable inlining
& \textbf{88.2\%}
& 0.0\%
& 29.4\%
& 29.4\%
& 70.6\%
& 52.9\%
& 64.7\%
& 70.6\% \\

\addlinespace
For-to-while lowering
& 81.0\%
& 9.5\%
& 85.0\%
& 14.3\%
& 85.7\%
& \textbf{100.0\%}
& 95.2\%
& 81.0\% \\

\bottomrule
\end{tabular*}

\vspace{3pt}
\begin{minipage}{0.98\textwidth}
\footnotesize
Scalabrino produced finite predictions for 97 of the 100 Java groups,
while all other methods processed all 100 groups. Percentages are
calculated over each method's available changed pairs.
\end{minipage}
\end{table*}

\begin{table*}[t]
\centering
\scriptsize
\setlength{\tabcolsep}{1.8pt}
\caption{Fine-grained response to independently applied readability
interferences in Python. DeepSeek denotes DeepSeek-V4-Pro-0813; GPT-6 and
GPT-6.1 denote GPT-6 Sol and GPT-6.1 Sol.}
\label{tab:interference-details-python}
\begin{tabular*}{\textwidth}{@{\extracolsep{\fill}}>{\raggedright\arraybackslash}p{0.27\textwidth}cccccccc@{}}
\toprule
\textbf{Interference}
& \textbf{Dorn}
& \textbf{Posnett}
& \shortstack{\textbf{Scala-}\\\textbf{brino}}
& \textbf{Mi}
& \textbf{DeepSeek}
& \textbf{GPT-6}
& \textbf{GPT-6.1}
& \textbf{\ourModel} \\
\midrule

Comment removal
& 78.3\%
& 40.5\%
& --
& 38.1\%
& 82.1\%
& 77.4\%
& 88.1\%
& \textbf{92.9\%} \\

\addlinespace
Identifier shortening
& 58.5\%
& 47.4\%
& --
& 27.4\%
& 88.4\%
& 84.2\%
& \textbf{92.6\%}
& 75.8\% \\

Misleading identifier substitution
& 30.5\%
& 46.9\%
& --
& 33.3\%
& 91.7\%
& 89.6\%
& \textbf{97.9\%}
& 85.4\% \\

Garbled identifier substitution
& 52.6\%
& 38.5\%
& --
& 27.1\%
& \textbf{96.9\%}
& 88.5\%
& 94.8\%
& 80.2\% \\

\addlinespace
Integer-literal encoding
& 44.8\%
& 13.3\%
& --
& 10.0\%
& 73.3\%
& 33.3\%
& 53.3\%
& \textbf{100.0\%} \\

Condition inversion and branch exchange
& 52.5\%
& 41.9\%
& --
& 11.3\%
& 95.2\%
& \textbf{98.4\%}
& 96.8\%
& 90.3\% \\

\addlinespace
Dead-branch injection
& 98.9\%
& 58.9\%
& --
& 24.2\%
& 96.8\%
& 96.8\%
& \textbf{100.0\%}
& 95.8\% \\

Readable unrelated-code injection
& 74.5\%
& 58.9\%
& --
& 32.6\%
& 83.2\%
& 89.5\%
& 92.6\%
& \textbf{96.8\%} \\

Low-readability unrelated-code injection
& \textbf{97.9\%}
& 58.9\%
& --
& 30.5\%
& 94.7\%
& 94.7\%
& 95.8\%
& \textbf{97.9\%} \\

\addlinespace
Partial layout compaction
& 75.6\%
& 56.0\%
& --
& 82.4\%
& 90.1\%
& 87.9\%
& \textbf{92.3\%}
& 85.7\% \\

\addlinespace
Numeric-intermediate introduction
& 55.6\%
& 25.0\%
& --
& 14.3\%
& 78.6\%
& \textbf{96.4\%}
& \textbf{96.4\%}
& 82.1\% \\

Intermediate-variable inlining
& \textbf{85.2\%}
& 3.6\%
& --
& 3.6\%
& 78.6\%
& 53.6\%
& 71.4\%
& 67.9\% \\

\addlinespace
For-to-while lowering
& 62.5\%
& 20.0\%
& --
& 16.0\%
& 96.0\%
& 96.0\%
& \textbf{100.0\%}
& 60.0\% \\

\bottomrule
\end{tabular*}

\vspace{3pt}
\begin{minipage}{0.98\textwidth}
\footnotesize
Scalabrino is omitted from the Python experiment because its released
implementation supports Java only. It produced finite predictions for
97 of the 100 Java groups. The reconstructed Dorn extractor processed
99 of the 100 Python groups. Percentages for these methods are calculated
over their available changed pairs.
\end{minipage}
\end{table*}

\clearpage

%% file: Appendices/ablation.tex
\section{Embedding-Only Model Results}
\label{app:embedding-only-results}

Before adding causal-LM predictability candidates, the same screening
procedure was applied to the 181 eligible traditional and
embedding-derived candidates. The resulting independently selected model
contains ten traditional and eight embedding-derived features. Its
selection procedure is reported in~\Cref{app:selection}.

The retained measurements are operator density, source-text entropy,
expression--literal density, longest line length,
computation-and-control pattern count,
identifier embedding dispersion, mean indentation change, structural-pattern
mixing under OPTICS and agglomerative clustering, structural embedding
dispersion, regular-expression prevalence, short-identifier prevalence,
short-identifier mathematical context, bitwise-operation prevalence,
unclustered structural-snippet ratio, identifier-length variation, decision
density, and operator-character share.

Performance of the embedding-only model under each of the five embedding
representations used during feature selection is reported
in~\Cref{tab:embedding-only-benchmarks}. The
same 18 feature definitions, Ridge pipeline, datasets, and folds are used
throughout; only the embedding-derived feature values change.

\begin{table}[H]
\centering
\small
\setlength{\tabcolsep}{4pt}
\caption{Performance of the independently selected 18-feature
embedding-only model across embedding representations. Each value is the
unweighted average of the six dataset-specific Spearman correlations.}
\label{tab:embedding-only-benchmarks}
\begin{tabular}{lrr}
\toprule
\textbf{Embedding model}
& \textbf{Pooled CV $\rho$}
& \textbf{LODO $\rho$} \\
\midrule
Nomic Embed Text v1.5 & \textbf{0.562} & 0.553 \\
Qwen3 Embedding 0.6B & \textbf{0.562} & 0.554 \\
Jina Embeddings v2 Base Code & 0.551 & 0.548 \\
Snowflake Arctic Embed M v2.0 & 0.551 & \textbf{0.561} \\
Voyage 4 Nano & \textbf{0.562} & 0.538 \\
\bottomrule
\end{tabular}
\end{table}

The pooled averages range from 0.551 to 0.562, and the LODO averages
range from 0.538 to 0.561. The embedding-only model therefore exhibits
similar aggregate performance across the five independently fitted
embedding representations.

%% file: Appendices/robustness.tex
\section{Model Robustness and Benchmark Uncertainty}
\label{app:model-robustness}

\subsection{Robustness Across Pretrained-Model Combinations}
\label{app:causal-lm-robustness}

We instantiate the fixed final feature schema using every pairing of the
three causal code language models and five embedding models used during
consensus selection. The 11 canonical feature definitions, Ridge pipeline,
six datasets, and fold assignments remain fixed, and each of the 15
combinations is fitted independently.

\begin{table}[H]
\centering
\scriptsize
\setlength{\tabcolsep}{5pt}
\caption{Performance across all 15 causal-LM--embedding-model
combinations. Values are unweighted means of the six dataset-specific
Spearman correlations in Panels A and B. Panels C and D report the
percentage of changed variants assigned a lower readability score than
their matched originals. Nomic, Qwen3, Jina, Arctic, and Voyage denote
the five embedding models.}
\label{tab:pretrained-model-grid}

\textbf{Panel A: Pooled 10-fold cross-validation}

\vspace{3pt}

\begin{tabular}{lrrrrr}
\toprule
\textbf{Causal language model}
& \textbf{Nomic}
& \textbf{Qwen3}
& \textbf{Jina}
& \textbf{Arctic}
& \textbf{Voyage} \\
\midrule
Qwen2.5-Coder-0.5B & 0.613 & 0.624 & 0.620 & 0.616 & 0.613 \\
DeepSeek-Coder-1.3B & 0.625 & 0.629 & 0.626 & \textbf{0.633} & 0.623 \\
OpenCoder-1.5B & 0.622 & 0.626 & 0.631 & 0.629 & 0.624 \\
\bottomrule
\end{tabular}

\vspace{0.8em}

\textbf{Panel B: Leave-one-dataset-out evaluation}

\vspace{3pt}

\begin{tabular}{lrrrrr}
\toprule
\textbf{Causal language model}
& \textbf{Nomic}
& \textbf{Qwen3}
& \textbf{Jina}
& \textbf{Arctic}
& \textbf{Voyage} \\
\midrule
Qwen2.5-Coder-0.5B & 0.624 & 0.622 & 0.625 & 0.628 & 0.617 \\
DeepSeek-Coder-1.3B & 0.621 & 0.629 & 0.631 & 0.633 & 0.625 \\
OpenCoder-1.5B & 0.630 & 0.629 & \textbf{0.638} & 0.637 & 0.627 \\
\bottomrule
\end{tabular}

\vspace{0.8em}

\textbf{Panel C: Java controlled-interference response}

\vspace{3pt}

\begin{tabular}{lrrrrr}
\toprule
\textbf{Causal language model}
& \textbf{Nomic}
& \textbf{Qwen3}
& \textbf{Jina}
& \textbf{Arctic}
& \textbf{Voyage} \\
\midrule
Qwen2.5-Coder-0.5B & \textbf{95.8\%} & 94.8\% & 95.1\% & 94.4\% & 95.6\% \\
DeepSeek-Coder-1.3B & 95.4\% & 95.3\% & 95.2\% & 94.9\% & 95.4\% \\
OpenCoder-1.5B & 94.8\% & 94.3\% & 93.8\% & 93.6\% & 95.2\% \\
\bottomrule
\end{tabular}

\vspace{0.8em}

\textbf{Panel D: Python controlled-interference response}

\vspace{3pt}

\begin{tabular}{lrrrrr}
\toprule
\textbf{Causal language model}
& \textbf{Nomic}
& \textbf{Qwen3}
& \textbf{Jina}
& \textbf{Arctic}
& \textbf{Voyage} \\
\midrule
Qwen2.5-Coder-0.5B & 85.8\% & 81.4\% & 84.7\% & 82.8\% & 84.8\% \\
DeepSeek-Coder-1.3B & 84.7\% & 79.7\% & 83.5\% & 80.7\% & 83.0\% \\
OpenCoder-1.5B & \textbf{88.0\%} & 84.2\% & 87.6\% & 85.5\% & 86.8\% \\
\bottomrule
\end{tabular}
\end{table}

All 15 independently fitted predictors retain the intended direction for
all 11 features.

\subsection{Bootstrap Confidence Intervals}
\label{app:benchmark-uncertainty}

We quantify uncertainty in the final OpenCoder and Jina-embedding
instantiation using a dataset-stratified nonparametric bootstrap. For each of
10,000 replicates, samples are drawn with replacement within each dataset,
Spearman correlation is recomputed for each dataset, and the average gives
equal weight to the six datasets. \ourModel's fitted parameters and out-of-fold
predictions remain fixed; the intervals therefore quantify sampling uncertainty
in the reported predictions rather than variation from retraining or feature
selection.

\begin{table}[H]
\centering
\small
\setlength{\tabcolsep}{5pt}
\caption{Spearman correlations with percentile-bootstrap 95\% confidence
intervals for \ourModel. The wider MBJP interval reflects its sample size of
17; the average intervals combine all six dataset-specific correlations with
equal weight.}
\label{tab:benchmark-bootstrap}
\begin{tabular}{lcc}
\toprule
\textbf{Dataset}
& \textbf{Pooled CV}
& \textbf{LODO} \\
\midrule
MBJP & 0.851 [0.512, 0.980] & 0.869 [0.557, 0.986] \\
Buse & 0.671 [0.541, 0.760] & 0.674 [0.551, 0.761] \\
Scalabrino & 0.549 [0.443, 0.643] & 0.560 [0.453, 0.649] \\
Dorn & 0.561 [0.477, 0.636] & 0.552 [0.467, 0.627] \\
Schnappinger & 0.625 [0.541, 0.697] & 0.633 [0.549, 0.704] \\
JetBrains & 0.531 [0.390, 0.647] & 0.540 [0.398, 0.660] \\
\midrule
Average & 0.631 [0.561, 0.673] & 0.638 [0.573, 0.679] \\
\bottomrule
\end{tabular}
\end{table}

We additionally compare the fixed \ourModel{} predictions with GPT-6.1 Sol
using a paired, dataset-stratified bootstrap. Each replicate draws the same
samples for both methods within each dataset and equally averages the six
resulting correlation differences.

\begin{table}[H]
\centering
\small
\setlength{\tabcolsep}{4pt}
\caption{Paired comparison with GPT-6.1 Sol over 10,000 bootstrap replicates.
The difference is \ourModel{} minus GPT-6.1 Sol in average Spearman
correlation.}
\label{tab:gpt61-paired-bootstrap}
\begin{tabular}{lrrrr}
\toprule
\textbf{Protocol}
& \textbf{\ourModel}
& \textbf{GPT-6.1}
& \textbf{Difference [95\% CI]}
& \textbf{$p$} \\
\midrule
Pooled CV & 0.631 & 0.537 & 0.094 [0.031, 0.157] & 0.0032 \\
LODO & 0.638 & 0.537 & 0.101 [0.038, 0.165] & 0.0008 \\
\bottomrule
\end{tabular}
\end{table}